\documentclass[acmsmall,screen]{acmart}

\usepackage{mathtools}
\usepackage{subcaption}
\usepackage{amsmath}
\usepackage{caption}
\usepackage{graphicx}
\usepackage{colortbl}
\usepackage{color}
\usepackage{braket}
\usepackage{hhline}
\usepackage{multirow}
\usepackage{braket}
\usepackage{makecell}

\newtheorem{definition}{Definition}

\newtheorem{example}{Example}

\definecolor{Light_Gray}{HTML}{EFEFEF}
\definecolor{Light_Orange}{HTML}{F9CB9C}
\definecolor{Light_Blue}{HTML}{9FC5E8}
\definecolor{Light_Green}{HTML}{D9EAD3}
\definecolor{Light_Yellow}{HTML}{FCE5CD}
\definecolor{Gray}{gray}{0.9}

\DeclarePairedDelimiter{\norm}{\lVert}{\rVert}
\begin{document}

\title{Using quantum annealing to generate test cases for cyber-physical systems}

\author{Hugo Araujo}
\email{hugo.araujo@kcl.ac.uk}
\affiliation{%
  \institution{King's College London}
  \country{United Kingdom}
}

\author{Xinyi Wang}
\email{xinyi@simula.no}
\affiliation{%
  \institution{Simula Research Laboratory}
  \country{Norway}
}

\author{Mohammad Reza Mousavi}
\email{mohammad.mousavi@kcl.ac.uk}
\affiliation{%
  \institution{King's College London}
  \country{United Kingdom}
}

\author{Shaukat Ali}
\email{shaukat@simula.no}
\affiliation{%
  \institution{Simula Research Laboratory}
  \country{Norway}
}

\begin{abstract}

Quantum computing has emerged as a powerful tool to efficiently solve computational challenges, particularly in simulation and optimisation. However, hardware limitations prevent quantum computers from achieving the full theoretical potential. Among the quantum algorithms, quantum annealing is a prime candidate to solve optimisation problems. This makes it a natural candidate for search-based software testing in the Cyber-Physical Systems (CPS) domain, which demands effective test cases due to their safety-critical nature. This work explores the use of quantum annealing to enhance test case generation for CPS through a mutation-based approach. We encode test case mutation as a binary optimisation problem, and use quantum annealing to identify and target critical regions of the test cases for improvement. Our approach mechanises this process into an algorithm that uses D-Wave's quantum annealer to find the solution. As a main contribution, we offer insights into how quantum annealing can advance software testing methodologies by empirically evaluating the correlation between problem size, hardware limitations, and the effectiveness of the results. Moreover, we compare the proposed method against state-of-the-art classical optimisation algorithms, targeting efficiency (time to generate test cases) and effectiveness (fault detection rates). Results indicate that quantum annealing enables faster test case generation while achieving comparable fault detection performance to state-of-the-art alternatives. 

\end{abstract}

\maketitle

\section{Introduction}

Quantum computing integrates quantum mechanics and computer science to change how complex computational problems can be analysed and solved. It takes advantage of quantum mechanics principles such as superposition and entanglement to analyse several possible solutions simultaneously, a phenomenon often referred to as \emph{quantum parallelism} \cite{nielsen2010quantum}. \emph{Superposition} means that \emph{qubits}, the quantum counterpart to classical bits, can represent a combination of values (typically represented as computational bases $\Ket{0}$ and $\Ket{1}$) simultaneously. Moreover, quantum computers can alter the state of multiple \emph{entangled} qubits simultaneously, even if only operating on one of them. Lastly, as a property that is particularly well-suited for optimisation problems, \emph{quantum tunnelling} allows particles to efficiently move towards the lowest energy state, which, in the optimisation context, represents the best solution. These three characteristics provide a powerful advantage to quantum computing, specifically when tackling simulation and optimisation problems. 

 
In particular, as one of the quantum algorithms that employs the use of such properties, quantum annealing~\cite{perdomo2015quantum} stands out as a promising method to solve optimisation problems. More specifically, the quantum annealers developed by D-Wave~\cite{cohen2014d} employ such principles to solve Quadratic Unconstrained Binary Optimization (QUBO) problems. It is designed to solve them by using a process where a large number of qubits are gradually transformed from a superposition of all possible states into a final state representing the solution.

In theory, quantum algorithms such as quantum annealing offer faster optimisation solutions than their state-of-the-art classical counterparts. In practice, however, the limited scale of current quantum computers restricts practical demonstrations of quantum advantage over classical systems. Hence, to efficiently take advantage of quantum computing, certain challenges need to be addressed. For instance, (i) encoding the problem in a way that can be solved by quantum computers and (ii) optimising qubits usage are amongst such challenges that warrant further research.

As an application domain that can take advantage of quantum computing, search-based software testing~\cite{matinnejad2015search, panichella2018automated, mateen2016optimization} is a prime candidate, as it involves optimising search techniques to generate test cases. This is particularly true in the subject domain of Cyber-Physical Systems (CPS), due to their intricate nature and large search space. Effective CPS testing
requires a robust test case generation strategy and one of its primary challenges lies in generating test suites that are both diverse and effective at detecting faults, without incurring prohibitive computational costs. Typical approaches either use efficient randomised methods, such as fuzzing~\cite{yang2024formal}, or effective search-based methods, via, for instance, optimisation solving~\cite{arrieta2017search}. Despite their utility, these strategies often present limitations in balancing testing effectiveness with computational efficiency due to the large problem search space.


As the main objective of this work, we aim to explore the use of quantum computing to efficiently generate test cases for CPS via a mutation-based, smart fuzzing approach. Mutation in the context of test case generation involves introducing changes or ``mutations'' into an existing set of test cases to create new ones. Instead of traditional, purely random mutations~\cite{silva2017systematic}, we employ a smart approach to identify when and how they are applied. Our approach uses quantum annealing to determine suitable regions of a test case to apply mutations by using metrics of testing adequacy to populate the objective formula. Then, we generate and embed mutations into the test cases to improve their effectiveness in terms of failure detection and improve their coverage in terms of diversity. We mechanise our approach into an algorithm that encodes the problem into a binary optimisation one and makes calls to the D-Wave's quantum annealer to solve it. 

We show the effectiveness (measured in fault detection rates) and efficiency (measured in time to generate test cases) of our approach by empirically evaluating the improved test suite against the ones generated by alternative, classical optimisation algorithms. In addition, we demonstrate the efficiency of our approach through a theoretical time complexity analysis. The results indicate that quantum annealing can generate test cases more efficiently than the alternatives and, therefore, as the main contribution of this work, we show that quantum annealing can generate test cases faster and with similar fault detection rates compared to the alternatives. 


\subsection{Research Questions}

The approach presented in this work is based on optimising CPS tests by applying mutations to inputs. In our context, such inputs are typically given as trajectories representing variable's valuations over time. Given the two-dimensional space (temporal and spatial) of the inputs, the choice of when and how to modify them has a crucial impact on the efficiency and effectiveness of the test mutation strategy. 

As a response to these challenges, this work embeds a quantum algorithm into a systematic process that identifies and mutates key areas of CPS test inputs. We hypothesise that applying quantum annealing to carefully select the regions of inputs to mutate has a significant impact on test case effectiveness. Additionally, we hypothesise that splitting the problem into sub-problems and using quantum annealing results in improved effectiveness compared to only using a classic state-of-the-art alternative. We turn our hypotheses into the following research questions.

\begin{itemize}

    \item \textbf{RQ1:} When using quantum annealing, what is the correlation between problem size, the number of physical qubits, and effectiveness of the results?

    \item \textbf{RQ2:} Is using quantum annealing to mutate the test cases more effective compared to the alternatives?

    \item \textbf{RQ3:} Is using quantum annealing to mutate the test cases more efficient compared to the alternatives, based on experimental results and theoretical analysis?
    

\end{itemize}

To answer these research questions, we evaluate our hypotheses via a controlled experiment that makes use of two CPS case studies (a platoon of connected vehicles and a pneumatic suspension system). These systems represent two distinct types of safety-critical CPS: the former operates over a layer of communication networks,  while the latter is strictly offline. The first research question relates to optimising the number of qubits and their correlation with the effectiveness of the results. The optimal values found via RQ1 are used to answer the remaining research questions. The second and third research questions relate to the use of quantum annealing and how it fares against the baselines.

\subsection{Paper Structure}

The remainder of this paper is organised as follows. In Section~\ref{sec:related}, we discuss the related work. In Section \ref{sec:background}, we describe the preliminary notations and techniques used in our work. In Section~\ref{sec:process}, we present our process on how to use quantum annealing to identify critical areas of a previously generated test suite and how to improve them by embedding mutations. In Section~\ref{sec:experiment}, we describe the design, execution, and results of our controlled experiment that aims to answer our research questions. Finally, in Section \ref{sec:conclusion}, we draw some conclusions and point out the directions of our future research.

\section{Related Work}
\label{sec:related}

Quantum computing research is rapidly advancing in various fields, such as finance~\cite{egger2020quantum}, drug discovery~\cite{blunt2022perspective}, and chemistry~\cite{motta2022emerging}. The application of quantum computing in software engineering is also emerging in recent years, especially in software testing~\cite{wang2024bootqa, wang2024quell, wang2024quantum}.  Industrial applications illustrate this trend, such as Bosch~\cite{bayerstadler2021industry} and BMW~\cite{bayerstadler2021industry}, who have reportedly used quantum computing for software testing. In addition, the elevator company Orona showcases a successful application of quantum extreme learning machines for regression testing~\cite{wang2024quell}. In research fields, Miranskyy applies Grover search to enhance the efficiency of dynamic testing on classical programs~\cite{miranskyy2022using}. 

More closely related to our work, quantum computing approaches are being investigated to be used in the context of test case optimisation. For instance, Hussein et al.~\cite{hussein2021quantum} uses amplitude amplification to solve test case minimisation problems via two quantum search algorithms. Wang et al.~\cite{wang2024quantum} used a QAOA-based algorithm (which aims to minimise a cost function via finding the lowest values for the energy function of a quantum system) to solve test case optimisation problems. Moreover, BootQA~\cite{wang2024bootqa} is a tool used to solve test case minimisation problems with quantum annealing. The tool employs a sampling technique to select test cases from a test suite, and feeds them to a quantum annealer through QUBO formulas. Unlike our work, none of the approaches mentioned above focus specifically on test case generation. More precisely, we employ quantum computing to identify the critical regions of test cases to be improved and, to our knowledge, no existing work in the literature employs a quantum algorithm in the generation of test cases.

To that end, however, many classical model-based test case generation approaches for CPS can be found in the literature~\cite{deshmukh2017testing, donze2010breach, arrieta2017search, menghi2020approximation, zhang2019uncertainty, araujo2019multiobjective, kim2019suggestion}. Sadri-Moshkenani et al.~\cite{sadri2022survey} published a comprehensive survey on state-of-the-art techniques used in this context. Several of these approaches employ a search-based heuristic to optimise a fitness function and guide the search to effective test cases, such as Bayesian optimisation~\cite{donze2010breach}, genetic algorithm~\cite{araujo2019multiobjective}, and NSGA-II~\cite{arrieta2017search}. Arrieta et al.~\cite{arrieta2018employing} utilises mutation operators and NSGA-II to mutate test cases and generate additional ones  while considering several criteria such as test case similarity and test execution time. Our work, instead,  specifically focuses on exploring the use of quantum annealing as an alternative to the classical heuristics used in many of the examples above. As a contribution to this work, in our controlled experiment, we provide empirical evidence that quantum annealing can be considered as a suitable alternative the classical counterparts.

Within the broader context of test case generation for a general class of systems, mutation-based fuzzing is a technique that aims to generate new test cases by mutating original ones, either randomly or with pre-defined mutation strategies~\cite{chen2018systematic, cha2012unleashing, chipounov2011s2e}. In certain contexts, steps can be taken to prevent test cases being rejected due to being invalid input sequences. For instance, one can set up constraints in the inputs space, such as rates of change in variable values or, in the context of code fuzzing, through grammar rules~\cite{pham2016model, kifetew2017generating}. Optimisation heuristics such as simulated annealing~\cite{bohme2017directed} and Markov Chains~\cite{bohme2016coverage} have also been used in the literature to optimise mutation-based fuzzers. As related work, Nichols et al. proposed a method for caster fuzzing~\cite{nichols2017faster} by incorporating Generative Adversarial Network (GAN) models. Böttinger~\cite{bottinger2018deep} formalised fuzzing as a reinforcement learning problem, where the fuzzing agent generates effective inputs by observing rewards caused by different mutations. We would like to reinforce that our work differs from the ones above by, instead of using classical algorithms, incorporating a quantum optimisation algorithm to mutate the test cases aiming for higher effectiveness and efficiency.



\section{Background}
\label{sec:background}

In this section, we discuss the preliminary knowledge used throughout this work. 

\subsection{Cyber-Physical Systems}

Cyber-physical systems (CPS) integrate computational systems into their physical environments. They represent a combination of discrete and continuous dynamics, often found in systems where a digital controller unit (discrete) is connected with some physical system (continuous). Particularly, a CPS might involve complex networks of these elements, also including human interaction. Examples of modern CPSs include vehicles and robotic systems~\cite{mosterman2016cyber}. Such systems require rigorous verification methodologies since they are usually safety-critical. The concept of a cyber-physical system is a generalisation of embedded systems, a collection of computing devices communicating with each other and interacting with the environment~\cite{alur2015principles}. 

As an example of a complex CPS, consider a car, driven by a human, that is followed by autonomous cars. The goal of the system is for the cars to maintain a comfortable distance from each other.  The computational systems are the controllers, the communication is dictated by protocols, and the control elements are the interactions with the physical world and the kinematics of the cars; for instance, switching gears also switches the acceleration profiles of these cars. The human factor is in the form of the leading driver. This system is called a platoon of connected vehicles and a model, along with its testing and analysis, is discussed as a case study in in Section~\ref{sec:case-studies}; in this same section, we discuss a second case study, which describes a vehicular suspension system that attempts to increase driver comfort.

Many classes or representations of cyber-physical systems can be found in the literature~\cite{de2009survey, alur2015principles}. In this work, we make use of Matlab/Simulink~\cite{documentationsimulation} - a widely-used development and simulation language for CPS - to implement models of CPS. The semantics of these models is represented by variable valuations and trajectories, formalised below:

\begin{definition}[Valuation]
\label{def:valuation}
Given a set of variables $V = \{X_1, \ldots, X_n\}$, we denote by $\mathit{Val}(V) = V \rightarrow \mathbb{R}$ the set of all total functions from $V$ to the real domain $\mathbb{R}$.
\end{definition} 

The notion of trajectory maps the moments in time to variable values.

\begin{definition}[Trajectory]
\label{def:trajectory}
Given a set of variables $V$, the set of trajectories over $V$,  denoted by $\mathit{Trajs}(V) = \{x_1, \ldots, x_m\}$, is the set of all mappings $T \rightarrow Val(V)$, where $T$ is the time domain, henceforth assumed to be a convex subset of $\mathbb{R}_{\geq 0}$.
\end{definition}

Essentially, trajectories describe a system's behaviour over time. In this work, both input and output variables are represented by trajectories. Typically, when testing CPS, a trajectory that comprises the set of input variables forms a test case and its verdict is given by performing checks on the corresponding output trajectory.

\subsection{Quantum Computing}

Quantum computing is a technology that applies principles of quantum mechanics to computation. Unlike classical computers, which use bits, quantum computing uses quantum bits (\textit{qubits}). 
There are two primary paradigms in quantum computing: \textit{gate-based quantum computing} and \textit{adiabatic quantum computing} (AQC). The former requires passing qubits through gates in a quantum circuit to perform calculations, the latter requires slowly changing the energy levels of qubits via, for example, laser pulses. The quantum annealing algorithm used in this work falls under the adiabatic quantum computing category.

\subsubsection{Adiabatic Quantum Computing}
AQC uses \textit{adiabatic theorem} (a theorem involving the energy levels of a system), to build a paradigm of quantum computing. The total energy of a quantum system can be described by a mathematical operator called a Hamiltonian, and its lowest possible energy is called the ground state. In AQC, an optimisation problem can be defined as a Hamiltonian, whose ground state encodes the optimal solution.

The complexity to determine the ground state of a complex Hamiltonian (e.g., Ising models~\cite{cipra1987introduction}) is NP-Complete~\cite{li2011solving}. 
Quantum computers can make use of the adiabatic theorem as a potential solution. This theorem states that if a quantum system is in the ground state of a Hamiltonian, and the Hamiltonian evolves `sufficiently slowly', the system will remain in the ground state throughout the process and end in the ground state of the final Hamiltonian. AQC implements this theorem into computation: a quantum system is first prepared in the ground state of an easy-to-solve, initial Hamiltonian. Then, the Hamiltonian evolves to a complex Hamiltonian, which encodes the problem meant to be solved and whose ground state encodes the desired solution. If the quantum system evolves slowly enough, it will remain in the ground state and end up at the ground state of the complex Hamiltonian, yielding the desired solution. 


However, in practice, it is difficult to maintain the condition in the adiabatic theorem due to the environment and the background noise. Additionally, changing a quantum system `sufficiently slowly' is not a trivial task and, as a result, the system might leave its ground state. Thus, quantum annealing is proposed as the relaxation of QAC, as it does not strictly adhere to the adiabatic theorem, making it a heuristic optimisation algorithm.

\subsubsection{Quantum Annealing}
\label{sec:quantum-annealing}

Quantum annealing is highly effective for tackling combinatorial optimisation problems, which aim to search through a large set of possible solutions to identify the optimal solution among all possibilities. It takes advantages of quantum properties (e.g., superposition, entanglement, and tunnelling) to explore solution spaces more efficiently than classical methods. The process of applying quantum annealing to solve optimisation problems is shown in Figure~\ref{fig:qa}. It comprises three main steps:

 \begin{figure}[!h]
        \centering
        \includegraphics[width=0.9\textwidth]{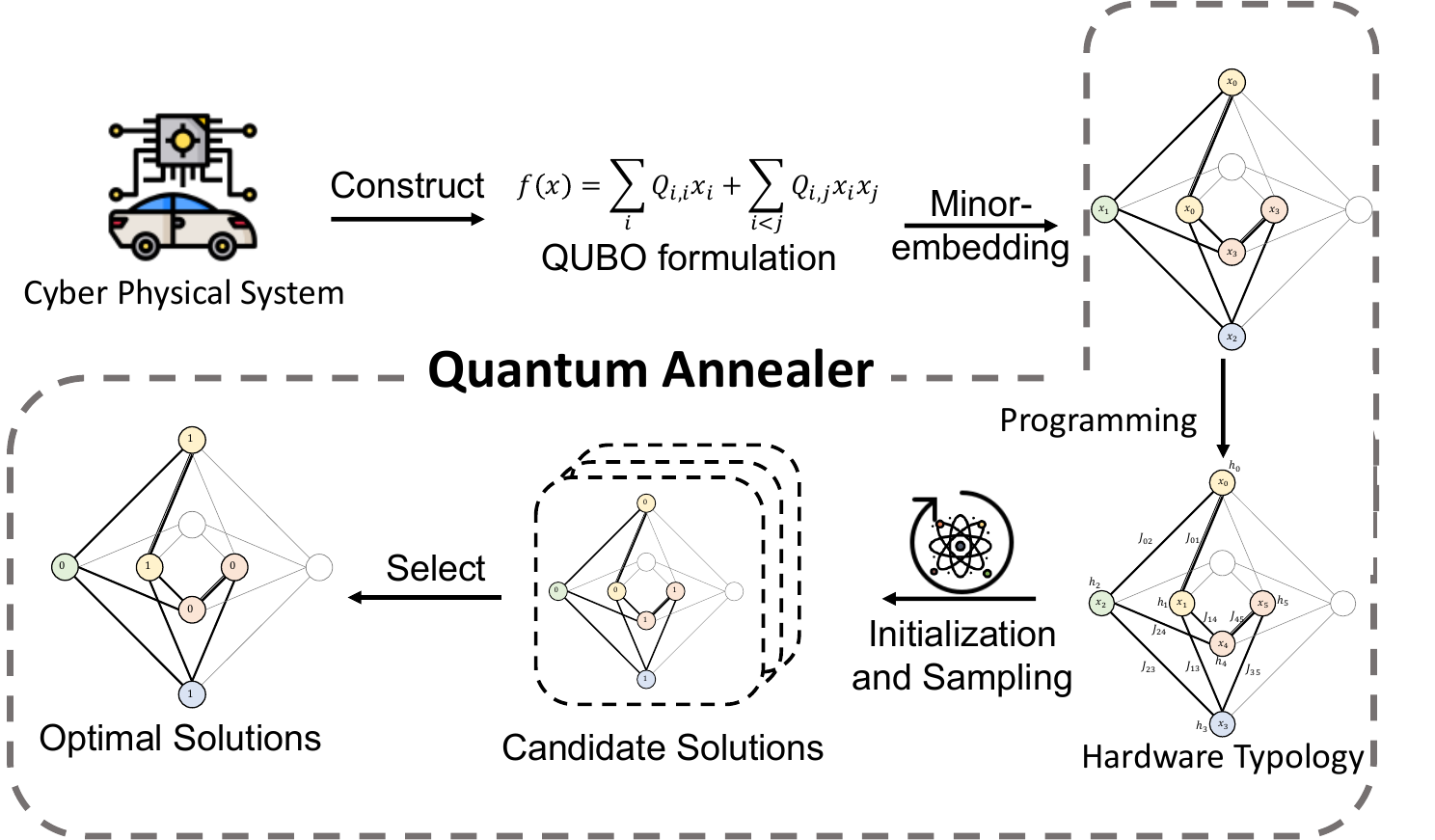}
        \caption{Quantum Annealing Process}
        \label{fig:qa}
    \end{figure}

\textbf{Step 1. QUBO construction:} To solve an optimisation problem using quantum annealing, the objective of the optimisation (defined by the domain experts) should be coded into a formula that can be mapped into the energy of a quantum system. For the D-Wave quantum computers, the final Hamiltonian is encoded as an Ising~\cite{cipra1987introduction} or a \textit{Quadratic Unconstrained Binary Optimisation (QUBO)}~\cite{punnen2022quadratic} model, which are mathematically equivalent. Therefore, the optimisation objective function should be formulated in one of these forms. In this paper, we use the QUBO formulation, defined below.

\begin{equation}\label{eq:qubo}
\min f(\boldsymbol{x}) = \boldsymbol{x}^{\mathrm{T}} Q \boldsymbol{x} = \sum_i Q_{i,i}x_i + \sum_{i<j} Q_{i,j}x_i x_j 
\end{equation} 

Similar to traditional optimisation algorithms, the QUBO formulation $f(\boldsymbol{x})$ (i.e., objective function) evaluates each candidate solution $\boldsymbol{x}$ in terms of fitness values, which describes the energy of a quantum system. Here $\boldsymbol{x}=[x_0, x_1, \ldots, x_{N-1}]$ is a vector consisting of binary decision variables $x_i$, and $\boldsymbol{x}^{\mathrm{T}}$ represents its transpose. The matrix $Q$ contains all coefficients in the objective function, whose values are determined by the specific optimisation problems, as defined by domain experts. The rightmost past of Equation~\ref{eq:qubo} represents the expanded equation. Considering the implementation of the quantum system, each decision variable $x_i$ represents a node of the quantum system (i.e, a qubit). Each qubit has a \textit{bias}, which is defined by the linear coefficients $Q_{i,i}$ of a corresponding variable. Moreover, qubits interact through couplers, where the two corresponding qubits are entangled, and the \textit{coupling strength} is decided by the quadratic coefficients $Q_{i,j}, (i<j)$ between two corresponding variables. 
Quantum annealing aims to find the optimal solution that can minimise the fitness value according to the $Q$ matrix.


\textbf{Step 2. Minor-embedding and programming:} The shape in which the physical qubits are physically or logically connected in hardware is called connection topology. In the process of \textit{programming} a QUBO formulation into the hardware, a QUBO formula is first mapped into a QUBO graph. During the programming step extra qubits may be required since current computers have limited connection topology. Hence, the original connection graph of the QUBO needs to be transformed into a new graph to fit the hardware topology by introducing \textit{qubit chains} in a process called \textit{minor-embedding}.

\begin{figure}[!tb]
\centering
\begin{subfigure}[b]{0.49\textwidth}
\centering
\includegraphics[width=0.49\textwidth]{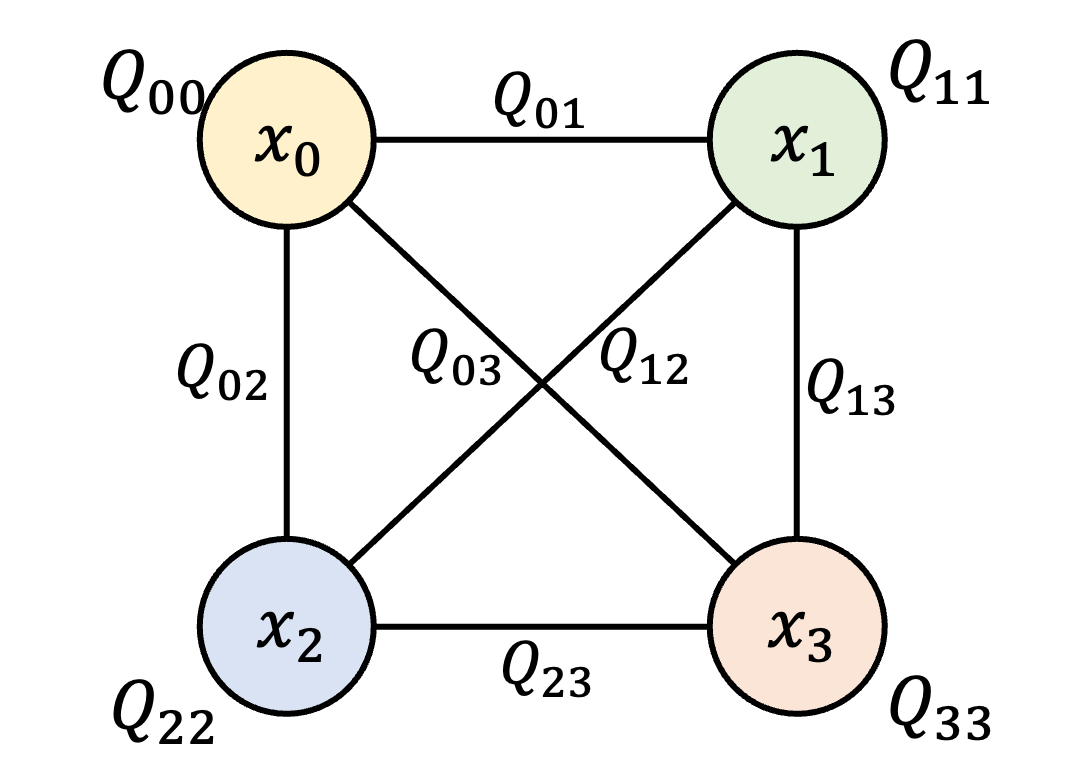}
\caption{QUBO graph}
\label{fig:QUBO_quboGraph}
\end{subfigure}
\hfill
\begin{subfigure}[b]{0.49\textwidth}
\centering
\includegraphics[width=0.49\textwidth]{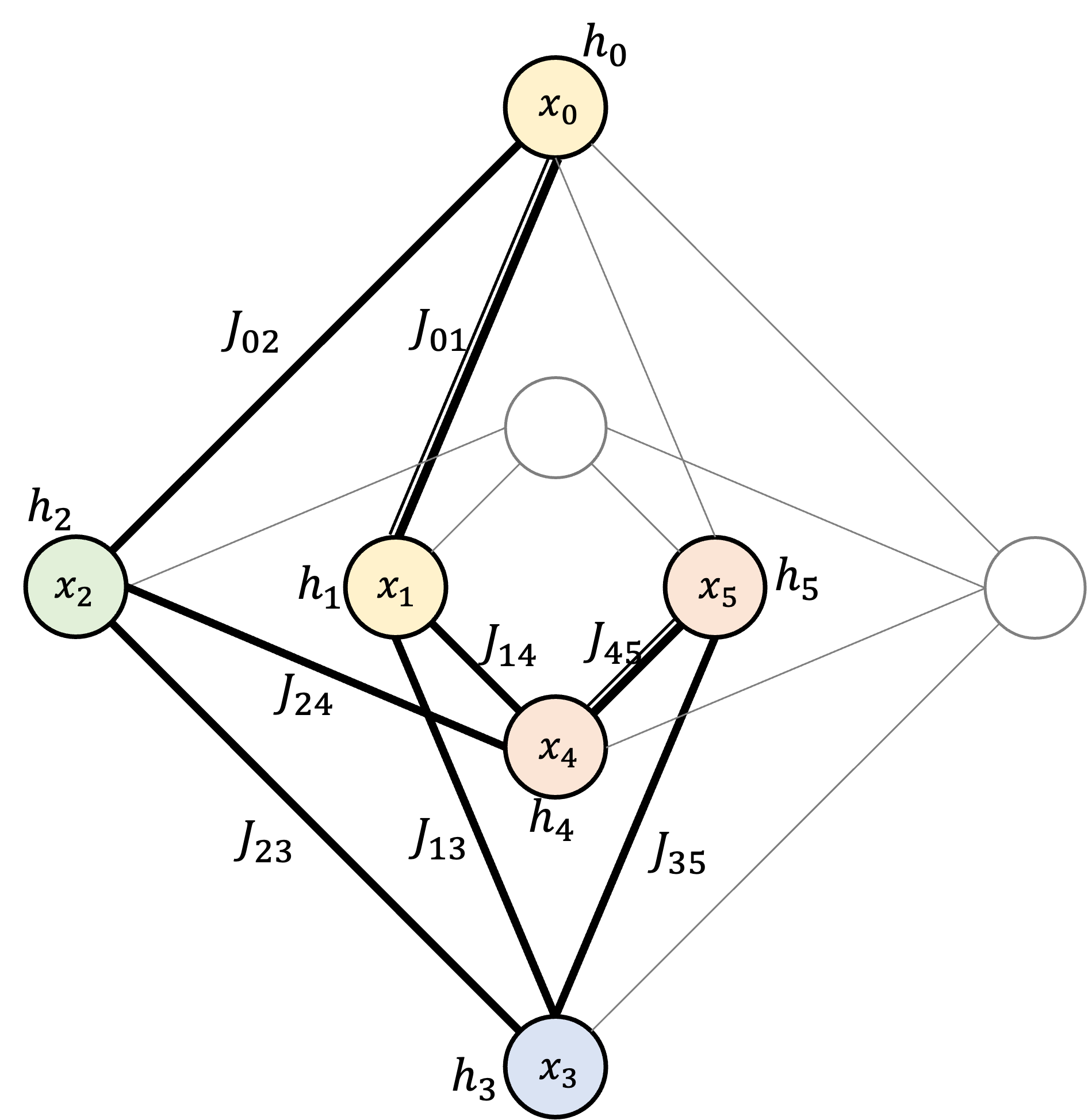}
\caption{QA hardware typology}
\label{fig:QUBO_workingGraph}
\end{subfigure}
\caption{QUBO formalisation for a four-variable problem}
\label{fig:QUBO}
\end{figure}

Consider the example in Figure~\ref{fig:QUBO}. Figure~\ref{fig:QUBO_quboGraph} shows a fully-connected QUBO graph with four variables, while Figure~\ref{fig:QUBO_workingGraph} depicts the topology supported by the D-Wave \textit{Advantage} system, where not all qubits are directly connected by couplers. Thus, in some cases, one variable should be mapped to a qubit chain (multiple connected qubits in the topology) to ensure any two pairs of the four variables are connected physically on the hardware. As shown in Figure~\ref{fig:QUBO_workingGraph}, $x_0$ and $x_3$ are both represented by a qubit chain of two qubits to remain connected to $x_1$ and $x_2$ as well as to each other. After the minor-embedding, in the programming step, all the coefficients of $Q_i$ and $Q_{i,j}$ will be assigned to the qubit bias and coupler.

\textbf{Step 3. Initialisation and sampling:} When the optimisation process begins, all qubits are first \textit{initialised} in the ground state of an easy-to-prepare Hamiltonian, where all qubits are in an equal superposition of all possible states. Then, the quantum system evolves from the initial Hamiltonian to the final Hamiltonian, which encodes the QUBO formulation of the optimisation problem to be solved, and its ground state encodes the solution. However, since it is difficult for quantum annealing to adhere to the adiabatic theorem, the output is not always guaranteed to be the exact solution. To minimise this uncertainty, multiple initialisation and sampling processes are iteratively performed, generating several \textit{candidate solutions}. The solution with the lowest energy is generally regarded as the \textit{optimal solution}.




\section{Quantum Annealing for test case Generation}
\label{sec:process}

In this section, we describe our methodology for generating test cases. We start with an overview of the process (Section~\ref{sec:overview}), followed by a detailed description of each of its steps: the generation of seed test cases and the collection of metrics (Section~\ref{sec:phase1}), solving the optimisation problem (Section~\ref{sec:phase2}), and, lastly, the mutation of test cases (Section~\ref{sec:phase3}).





\subsection{Running Example}
\label{sec:running}

\begin{figure}[!h]
        \centering
        \includegraphics[width=0.7\textwidth]{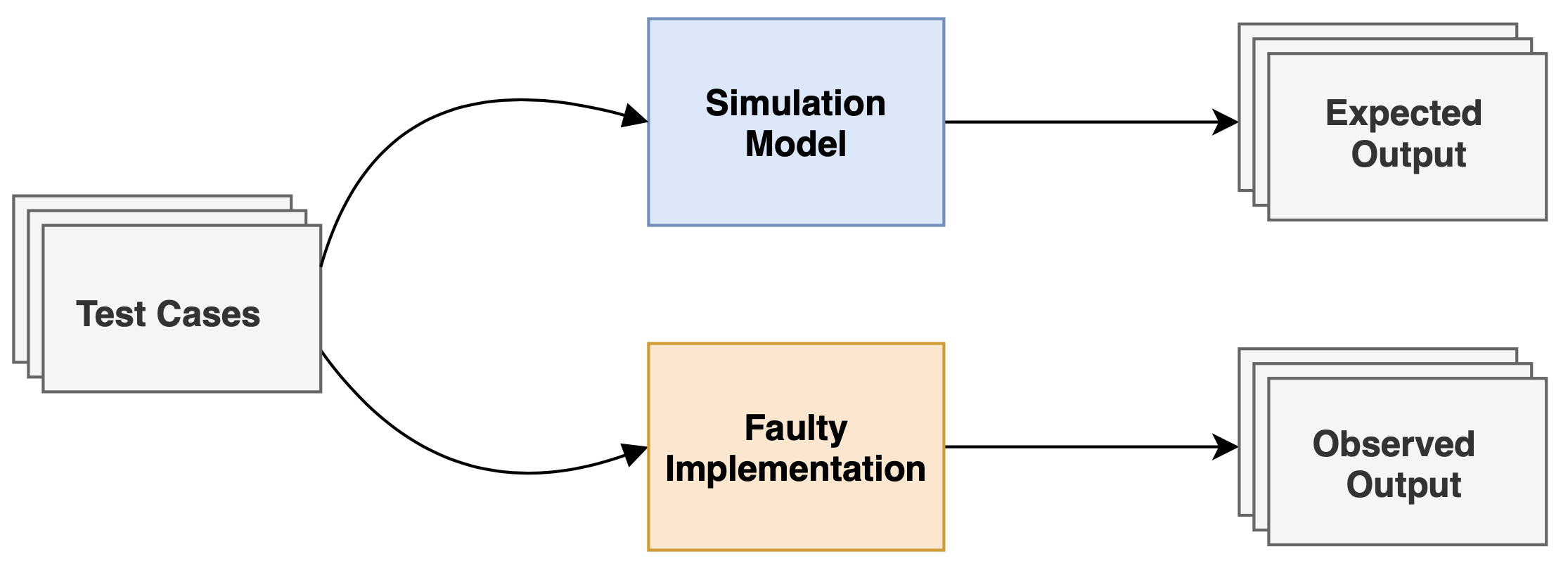}
        \caption{Testing the running example.}
        \label{fig:running-example}
    \end{figure}

Consider as our running example a CPS that describes a minimalistic vehicle engine that receives the position of the accelerator pedal as input (valued between 0 and 1) and outputs the amount of fuel (valued between 0mL and 5mL) to be injected into cylinders. For simplicity, this system is designed to directly map the position of the pedal to the amount of fuel. For instance, when the pedal is on 0, the amount of fuel to be injected should be 0 mL. Conversely, when the pedal is on 1 the amount of fuel released should be 5 mL. Assume that the test engineers of this system have created a simulation test model that faithfully captures the intended behaviour. However, a hardware failure in the physical implementation resulted in the engine releasing more fuel than it should (by 0.2 mL). To test this system, test engineers generate test cases and compare the output of the simulation model (expected output) against that of the system implementation (observed output), as depicted in Figure~\ref{fig:running-example}.

In what follows, we describe how we use quantum annealing to generate test cases that can make the failure more evident and, therefore, more easy to detect. 


\subsection{Overview of the process}
\label{sec:overview}

\begin{figure}[!h]
        \centering
        \includegraphics[width=1\textwidth]{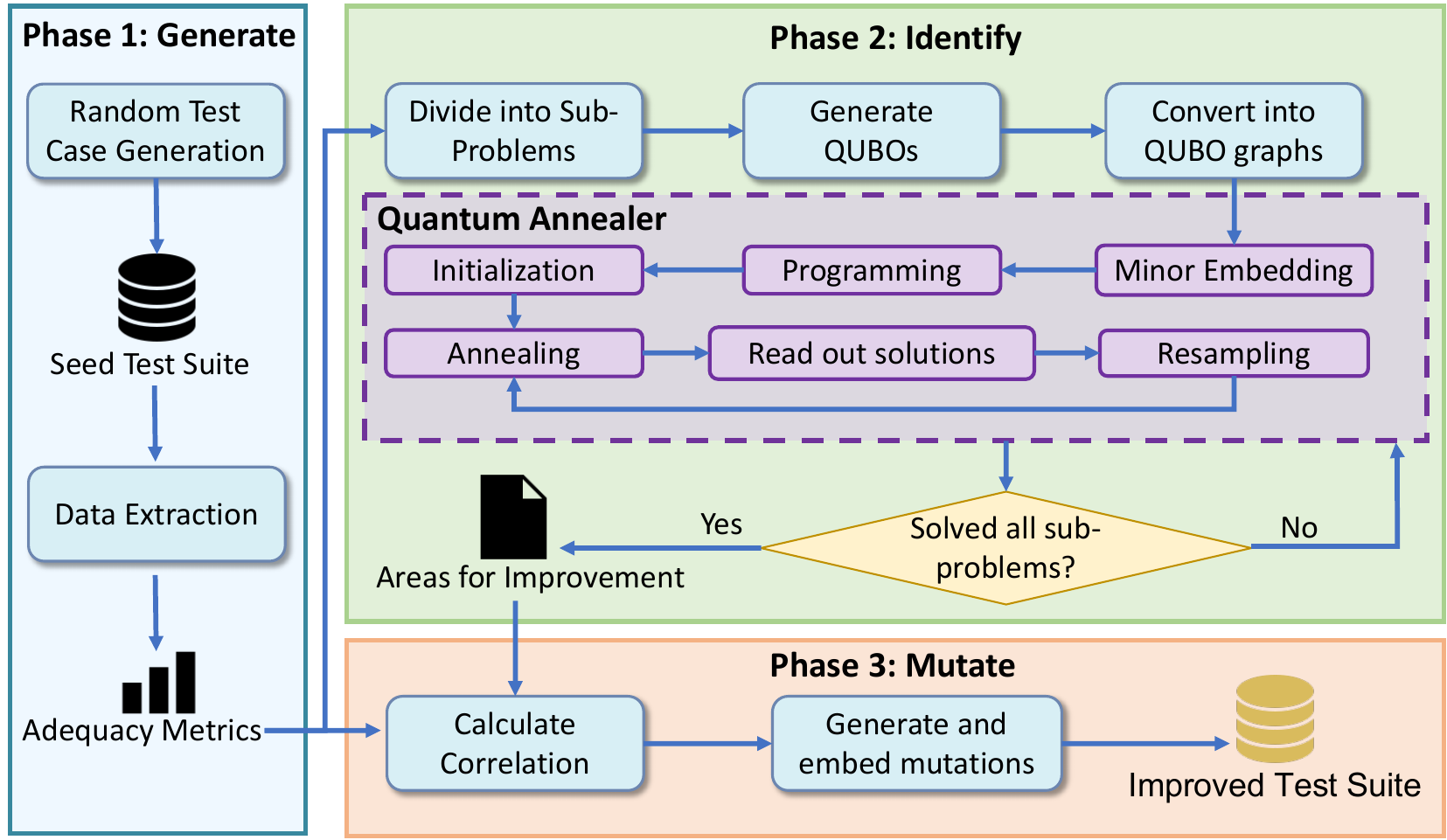}
        \caption{Overview of the process.}
        \label{fig:process}
    \end{figure}

An overview of our test case generation process is depicted in Figure~\ref{fig:process}. The goal is to modify test cases by identifying key areas and mutating them to improve their effectiveness. In this work, a test case is represented by an input trajectory, which, in practice, can be seen as a time-series. The process comprises three phases. In Phase~1, a suite of seed for test cases are randomly generated and executed on the SUT and, thus, yielding outputs. Then, we collect a series of metrics (related to the adequacy of the test suite) from both the test cases and their respective outputs. In Phase~2, these adequacy metrics are used to identify critical areas of the test cases. We formulate an optimisation problem that attempts to maximise or minimise our metrics and we solve this problem using a quantum annealing algorithm. In Phase~3, once the critical areas in each test case have been identified, we improve them by using an algorithm that generates the mutations and embeds them into the original test cases.

\subsection{Phase~1: Generate}
\label{sec:phase1}
In Phase~1, random values are generated for data points along each test case, i.e., trajectory over time ($t \in T$), where $T$ refers to the test duration. These values are constrained by minimum and maximum boundaries, as well as a maximum allowable rate of increase or decrease to ensure that the test case is valid. 
To form a smooth trajectory, these data points are connected via a piecewise non-linear curve fitting solution provided by the Matlab Optimisation Toolbox~\footnote{https://www.mathworks.com/products/optimization.html}. We follow this approach for all input variables to form the trajectory that represents a test case. A set of test cases forms a test suite. The test suite is then executed on the SUT to generate the output for the corresponding test case.

\begin{example}

Consider our running example. Figure~\ref{fig:test-suite} depicts three test cases in an initial test suite and the respective outputs (expected and observed). Each test case, in practice, is essentially a series of values that, when connected, form a trajectory.

    \begin{figure}[!h]
  \centering
    \begin{subfigure}{0.32\linewidth}
      \centering
      \includegraphics[width=1\linewidth]{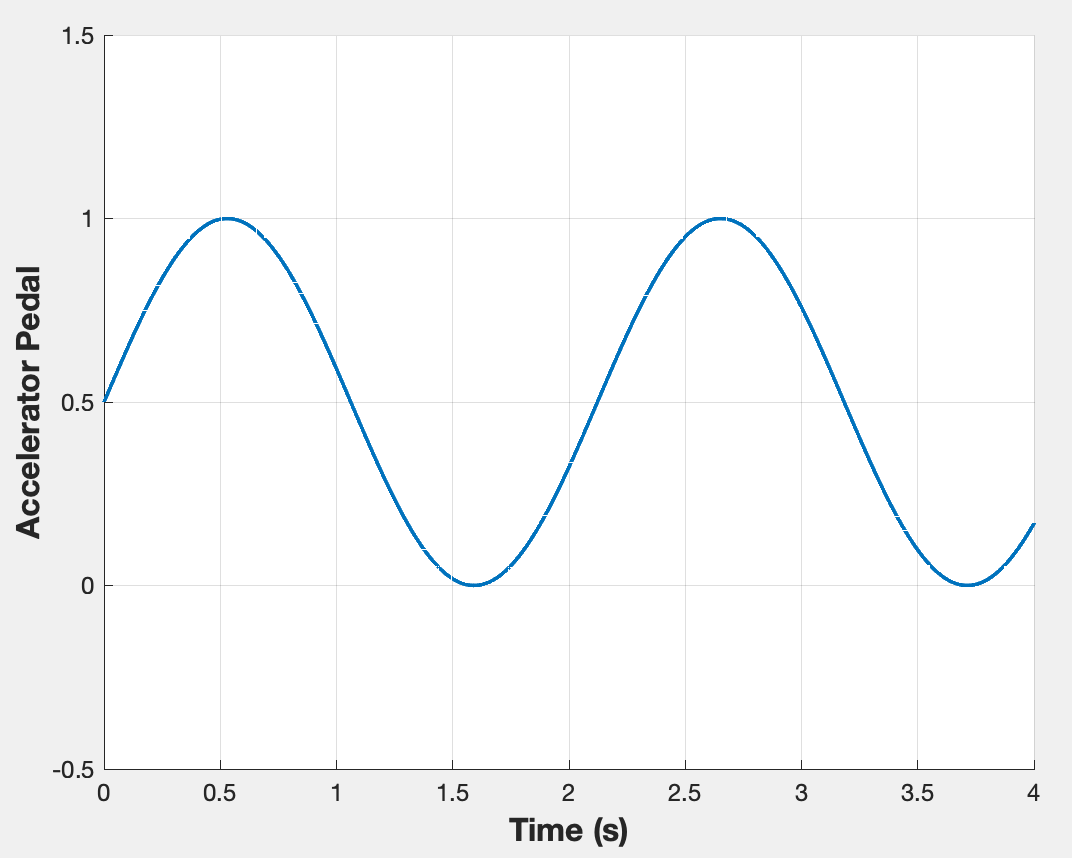} 
      \caption{Test case 1.}
      \label{fig:re-in1}
    \end{subfigure} \hfill
    \begin{subfigure}{0.32\linewidth}
      \centering
      \includegraphics[width=1\linewidth]{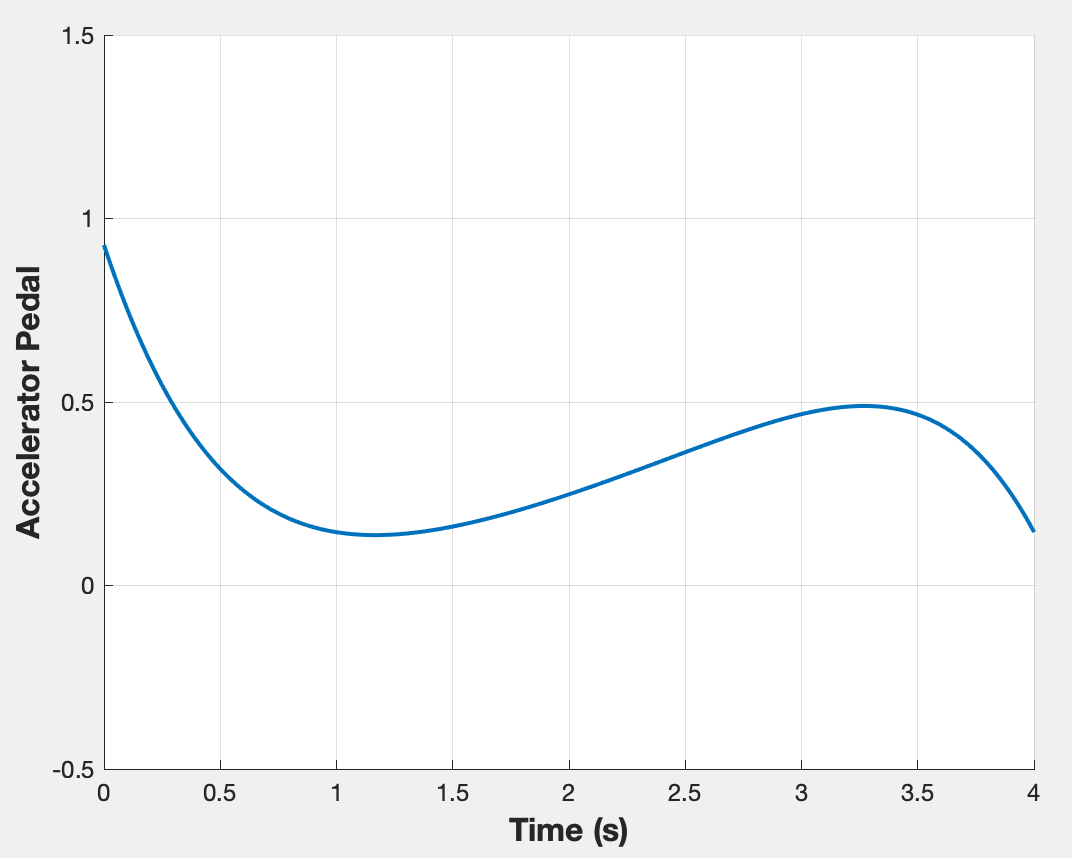} 
      \caption{Test Case 2.}
      \label{fig:re-in2}
    \end{subfigure} \hfill
    \begin{subfigure}{0.32\linewidth}
      \centering
      \includegraphics[width=1\linewidth]{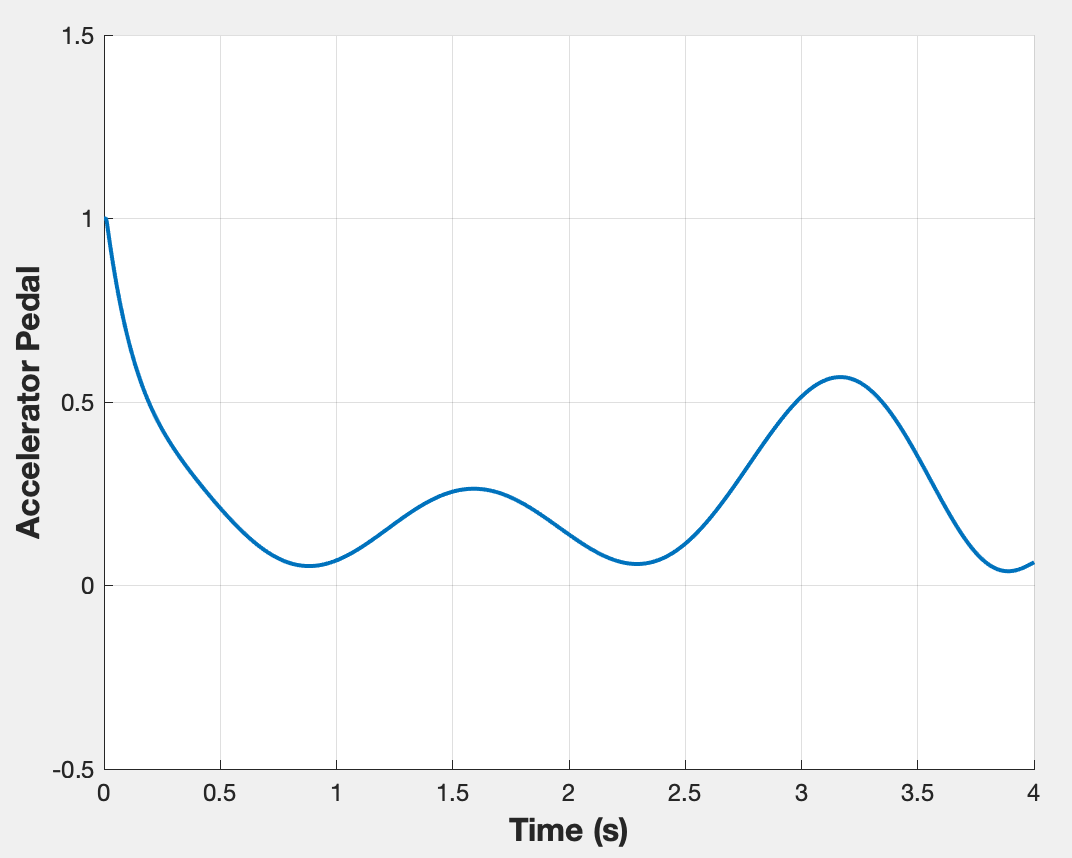} 
      \caption{Test Case 3.}
      \label{fig:re-in3}
    \end{subfigure}\\
    \begin{subfigure}{0.32\linewidth}
      \centering
      \includegraphics[width=1\linewidth]{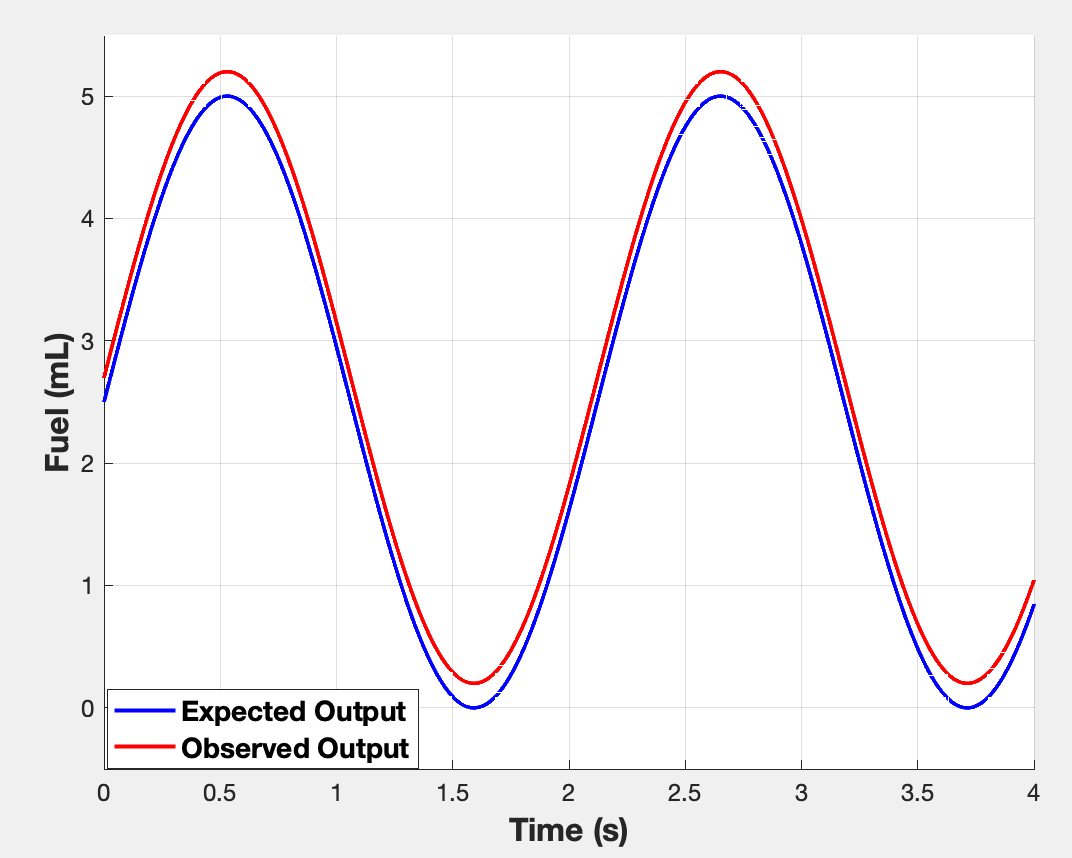} 
      \caption{Outputs from Test Case 1.}
      \label{fig:re-ot1}
    \end{subfigure} \hfill
    \begin{subfigure}{0.32\linewidth}
      \centering
      \includegraphics[width=1\linewidth]{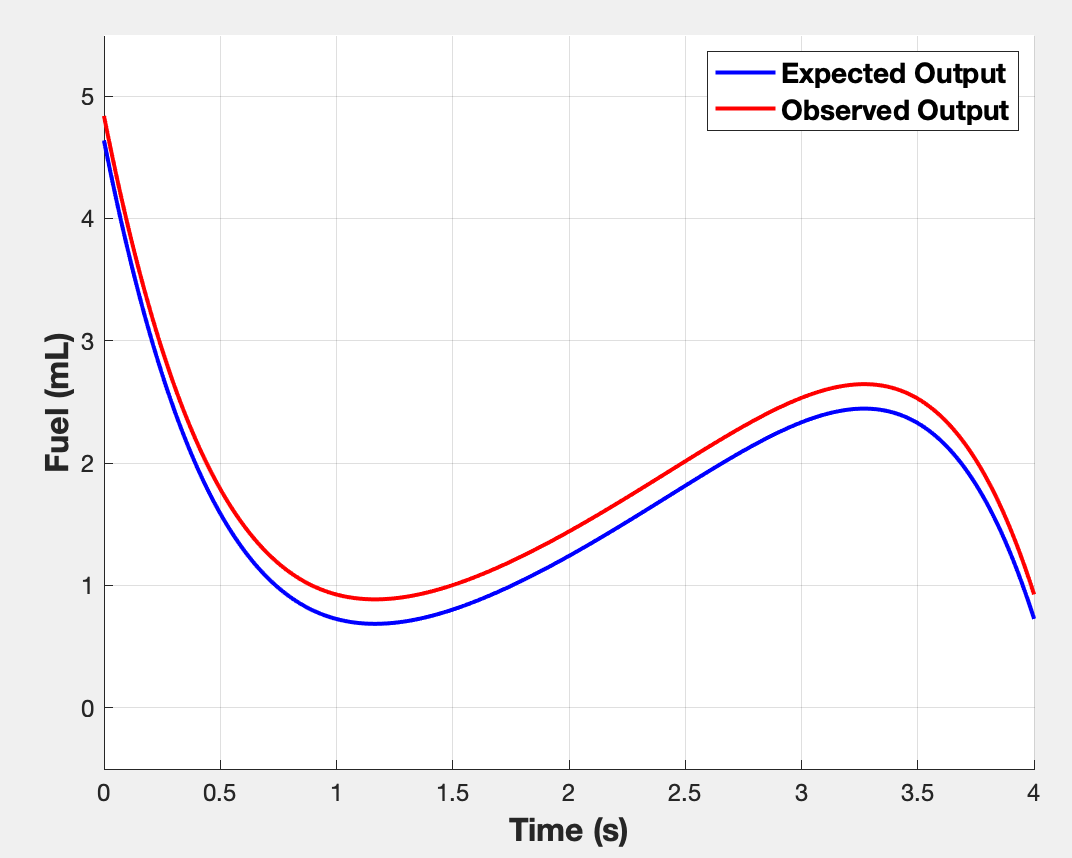} 
      \caption{Outputs from Test Case 2.}
      \label{fig:re-ot2}
    \end{subfigure} \hfill
    \begin{subfigure}{0.32\linewidth}
      \centering
      \includegraphics[width=1\linewidth]{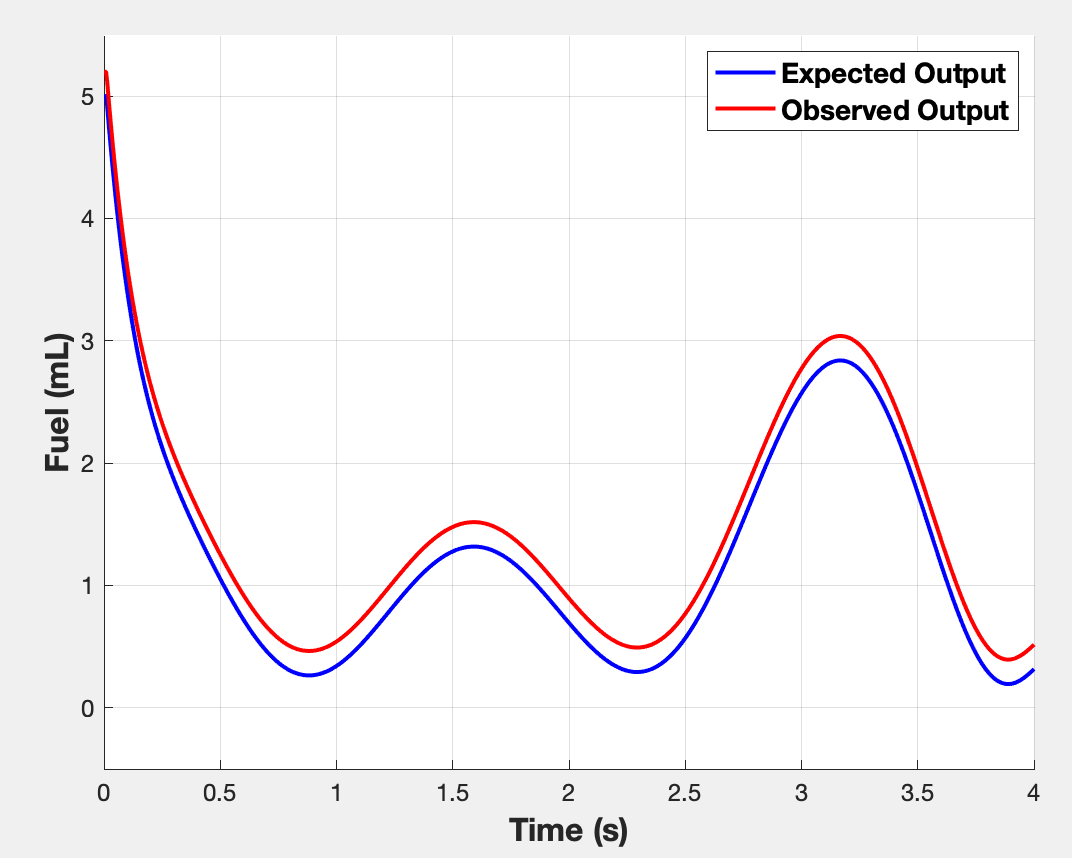} 
      \caption{Outputs from Test Case 3.}
      \label{fig:re-ot3}
    \end{subfigure}
    \caption{Test cases in the test suite.}
  \label{fig:test-suite}
\end{figure}
    
\end{example}

For each data point in a test case (i.e., the value of a point in time), three adequacy metrics are collected: one related to effectiveness and two related to diversity (i.e., input and output diversity) of the trajectory. 

\textit{Effectiveness metric ($\mathcal{E}$):} We calculate the Euclidean distance between observed output $O(t)$ (i.e., obtained from exercising the input directly on the SUT) and the expected output $E(t)$ (i.e., observed from exercising the input on a design model) at time $t$. The metric value for each data point $I(t)$ is then normalized using the min-max normalisation method.

    \[ \mathcal{E}(I(t)) = \frac{\norm{O(t) - E(t)} - \mathcal{R}_{min}}{\mathcal{R}_{max}-\mathcal{R}_{min}} \]

In this formula, $\mathcal{R}_{max}$ and $\mathcal{R}_{min}$ represent the maximum and minimum values for the range of the signal.

\textit{Input diversity ($\mathcal{D}_i$) and output diversity ($\mathcal{D}_o$): } Similarly, we calculate the input and output diversity for each data point in the test case. For each data point at a time $t$, we select a fixed range of surrounding data points (e.g., all data points within the $t \pm r$ range).
This reduces the original trajectory to a slice of it ($s_1$). The diversity metric is then calculated by using normalised Euclidean distance between points of the input slice ($s_1$) and slices (within the same $[t-r,t+r]$ range ) in the test suite ($s_i$). Similar to the effectiveness metric, the diversity formula normalises the diversity between the $[\mathcal{R}_{min},\mathcal{R}_{max}]$ range. This is given by the formula below, which was proposed by Matinnejad et al.~\cite{matinnejad2015effective}.

\[ \mathcal{D}_i (t) = \frac{\sqrt{\sum_{w=(t-r)}^{t+r}(s_1(w)-s_i(w))^2}}{\sqrt{(t+r)+1} \times (\mathcal{R}_{max} - \mathcal{R}_{min}) } \]



The input diversity for a data point is given by the minimum distance between the respective slice and the corresponding slices in the other test cases. To calculate the output diversity of a test case data point, we take a slice comprising the surrounding points within a $\pm\ t$ range from the output of that test case. Then, we use the same formula above to compare the slice of the respective output against slices of the same time interval for every other output. The diversity is determined by calculating the minimum distance, analogous to how we calculate input diversity.

\begin{example}

Consider our running example. Table~\ref{fig:test-suite} shows the values for the three metrics that were computed when running the three test case. Here, we only show the values for the first 10 data points; the time difference between each data point is 0.01s.

\begin{table}[!h]
\caption{Metrics for each test case in the running example}
\label{tab:metrics}
\begin{tabular}{cl|ccc|l|ccc|l|ccc|}
 \hhline{~~---~---~---}
 &  & \multicolumn{3}{c|}{\cellcolor[HTML]{EFEFEF}\textbf{Test Case1}} &  & \multicolumn{3}{c|}{\cellcolor[HTML]{EFEFEF}\textbf{Test Case 2}} &  & \multicolumn{3}{c|}{\cellcolor[HTML]{EFEFEF}\textbf{Test Case 3}} \\ 
  \hhline{~~---~---~---}
\multirow{-2}{*}{\textbf{Index}} &  & \multicolumn{1}{l}{\cellcolor[HTML]{EFEFEF}\textbf{Efctvt.}} & \multicolumn{1}{l}{\cellcolor[HTML]{EFEFEF}\textbf{I. Div.}} & \multicolumn{1}{l|}{\cellcolor[HTML]{EFEFEF}\textbf{O. Div.}} &  & \multicolumn{1}{l}{\cellcolor[HTML]{EFEFEF}\textbf{Efctv.}} & \multicolumn{1}{l}{\cellcolor[HTML]{EFEFEF}\textbf{I. Div.}} & \multicolumn{1}{l|}{\cellcolor[HTML]{EFEFEF}\textbf{O. Div.}} &  & \multicolumn{1}{l}{\cellcolor[HTML]{EFEFEF}\textbf{Efctv.}} & \multicolumn{1}{l}{\cellcolor[HTML]{EFEFEF}\textbf{I. Div.}} & \multicolumn{1}{l|}{\cellcolor[HTML]{EFEFEF}\textbf{O. Div.}} \\ \cline{3-5} \cline{7-9} \cline{11-13} 
1 &  & 0.305 & 0.697 & 0.956 &  & 0.417 & 0.100 & 0.183 &  & 0.126 & 0.148 & 0.240 \\
2 &  & 0.308 & 0.696 & 0.956 &  & 0.413 & 0.097 & 0.183 &  & 0.123 & 0.148 & 0.240 \\
3 &  & 0.310 & 0.696 & 0.956 &  & 0.411 & 0.097 & 0.183 &  & 0.122 & 0.148 & 0.240 \\
4 &  & 0.312 & 0.695 & 0.956 &  & 0.406 & 0.097 & 0.183 &  & 0.120 & 0.145 & 0.240 \\
5 &  & 0.316 & 0.695 & 0.956 &  & 0.401 & 0.097 & 0.183 &  & 0.117 & 0.145 & 0.240 \\
6 &  & 0.317 & 0.695 & 0.956 &  & 0.400 & 0.097 & 0.183 &  & 0.113 & 0.145 & 0.240 \\
7 &  & 0.320 & 0.696 & 0.956 &  & 0.399 & 0.097 & 0.183 &  & 0.112 & 0.145 & 0.240 \\
8 &  & 0.321 & 0.696 & 0.956 &  & 0.398 & 0.097 & 0.183 &  & 0.111 & 0.145 & 0.240 \\
9 &  & 0.324 & 0.696 & 0.956 &  & 0.394 & 0.097 & 0.183 &  & 0.111 & 0.145 & 0.240 \\
10 &  & 0.324 & 0.696 & 0.956 &  & 0.393 & 0.096 & 0.183 &  & 0.107 & 0.145 & 0.240 \\ \cline{3-5} \cline{7-9} \cline{11-13} 
\end{tabular}
\end{table}

\end{example}

\subsection{Phase~2: Identify}
\label{sec:phase2}


In this phase, we aim to identify the data points in each test case to improve the corresponding critical areas. We first decompose the problem into sub-problems and then solve each of them using quantum annealing. The sub-solutions are merged at the end to obtain the final solution.

\emph{\textbf{Dividing the problem into sub-problems:}} A QUBO problem can be visualised as a QUBO graph, which requires a large number of qubits and couplers to be fully represented. Since current quantum annealing hardware has a limit on the number of qubits, a workaround is needed to solve large problems: a sampling method breaks the problem into smaller sub-problems. This technique involves randomly selecting subsets of data, repeatedly, to create samples representing the entire dataset. This strategy generates slices of trajectories from the full input trajectory (i.e., the test case), ensuring each smaller sub-problem (i.e., a slice) can be solved using current hardware. 

More specifically, given a test case, we first divide it into $M$ smaller slices of the same fixed range. From each slice, we sample $N$ data-point, where $N$ is based on the number of qubits and couplers the hardware can support. Hence, we divide the problem into $M$ subsets of size $N$, which are randomly sampled from the full trajectory $i$ until a specified percentage is covered. These $M$ subsets are used to formulate $M$ sub-problems based on the QUBO formulation, and each sub-problem is solved separately using QA, resulting in the corresponding $M$ sub-solutions. Eventually, the data-points selected in each sub-solution are combined to form the final solution.

\emph{\textbf{Identifying critical areas with Quantum Annealing:}} For each sub-problem generated in the previous steps, we construct an overall objective function in the QUBO formulation (introduced in Section~\ref{sec:background}) to select the data points. We represent the selection of each data point as a binary variable $x_i$ in the QUBO. We set 1 to $x_i$ if the $i$-th data point is selected; otherwise, we set it to 0. 

This objective function integrates four objectives and a constraint into a QUBO model, using a weighted-sum strategy~\cite{wang2023test}. Specifically, three of the objectives relate to the previously introduced metrics 
The fourth objective is related to avoiding selecting too many data points, thus aiming to minimise the number of selected data points.
The constraint is applied to control the distance among the selected data points.

To build these three objectives, we first extract the effectiveness, input, and output diversity values of each data point in the sub-problem. Then, we minimise the distance between the sum of the metric values of the selected data points and the corresponding ideal value for that metric~\cite{wang2023test}. In detail, since we would like to maximise the sum of effectiveness values, we calculate the upper limit $L^{\mathsf{ef}}$, which is the total sum of the effectiveness values for all data points. On the other hand, the minimal  input and output diversity values are $L^{\mathsf{id}}=0$ and $L^{\mathsf{od}}=0$. We represent the $\mathsf{mr}\in \{\mathsf{ef, id, od}\}$ as the generalised metric. For each metric, the metric value of each data point is represented as $v_i^{\mathsf{mr}}$. Given a trajectory with a total of $N$ data points, the selection of all data points can be represented in a vector $\boldsymbol{x}=[x_0, x_1, \ldots, x_{N-1}]$. 

We represent the objective function as
\begin{equation}
    \min \mathit{Obj}^{\mathsf{mr}}(\boldsymbol{x}) = (\sum_{i=0}^{N-1}v^{\mathsf{mr}}_i x_i-L^{\mathsf{mr}})^2, \mathsf{mr} \in \{\mathsf{ef, id, od}
\end{equation}

Since $x_i \in \{0,1\}$, we have that $x_i^2=x_i$. Meanwhile, we do not consider the constant terms in QUBO since they do not affect the optimisation. Thus, we can expand the objective as follows:
\begin{equation}
    \label{eq:objective-function-1}
    \min O^{\mathsf{mr}}(\boldsymbol{x}) = \sum_{i=0}^{N-1}((v_i^{\mathsf{mr}})^2-2L^{\mathsf{mr}}v_i^{\mathsf{mr}})x_i+2\sum_{i<j}^{N-1}v_i^{\mathsf{mr}}v_j^{\mathsf{mr}}x_ix_j, \mathsf{mr} \in \{\mathsf{ef, id, od}\}
\end{equation}

Mapping to the QUBO formulation, $(v_i^{\mathsf{mr}})^2-2L^{\mathsf{mr}}v_i^{\mathsf{mr}}$ are the linear coefficients of the variable $x_i$ while $2v_i^{\mathsf{mr}}v_j^{\mathsf{mr}}$ are the quadratic coefficients between variables $x_i$ and $x_j$.

\begin{example}
    Suppose that the sampling selected two data points (N = 2) from Test Case 1. The index of the selected points are 3 and 7. The vector of binary variables ($x$), can have four possible values $([0,0],[0,1],[1,0],[1,1])$, where, for instance, $x = [1,0]$ means that the data point of index 3 is selected but the data point of index 7 is not.
    For $x = [1,1]$ and using effectiveness metric ($ef$) as the the only objective, replacing the values in Equation~\ref{eq:objective-function-1} leads to the following:

    \begin{align*}
        \mathit{O}^{\mathsf{ef}}([1,1]) &= \sum_{i=0}^{2-1}((v_i^{\mathsf{ef}})^2-2L^{\mathsf{ef}}v_i^{\mathsf{ef}})x_i+2\sum_{i<j}^{N-1}v_i^{\mathsf{ef}}v_j^{\mathsf{ef}}x_ix_j\\
        &= (.31^{2}-2*.63*.31*1) + (.32^2-2*.63*.32*1)+ 2 * (.31*.32*1*1)\\
        &= -0.398
    \end{align*}

    On the other hand, $O^{\mathsf{ef}}([0,0]) = 0$, $O^{\mathsf{ef}}([1,0])= -0.294$, and $O^{\mathsf{ef}}([0,1]) = -0.3$, which indicates that, when it comes to effectiveness only, selecting both data points results in the optimal value.
\end{example}

In the fourth objective, similarly, we minimise the distance between the total number of selected data points in the sub-problem and the theoretical lower limit $L^{\mathsf{num}}=0$. We create one more objective:
\begin{equation}
    \min \mathit{Obj}^{\mathsf{num}}(\boldsymbol{x}) = (\sum_{i=0}^{N-1} x_i - L^{\mathsf{num}})^2
\end{equation}

We can get the linear and quadratic coefficients of corresponding variables in the QUBO formulation similar to the above. With the four objectives, we can combine them into an overall QUBO formulation with the weighted-sum method, which gives us:

\begin{equation}
    \label{eq:objective-function-2}
    \min O(\boldsymbol{x}) =\sum_{\mathsf{mr} \in \{\mathsf{ef, id, od}\}}w_{\mathsf{mr}}O^{\mathsf{mr}}(\boldsymbol{x})+w_{\mathsf{num}}O^{\mathsf{num}}
\end{equation}

\begin{example}
    Consider that the sampling selected three data points (N = 3) from Test Case 2. The index of the selected points are 1, 4, and 7. For this example, we consider all possible objectives ($\mathsf{ef, id, od, num}$) and the weights are set to $w_{\mathsf{ef}} = 0.25, w_{\mathsf{od}} = 0.125, w_{\mathsf{id}} = 0.125, w_{\mathsf{num}} = 0.5$. The vector of binary variables ($x$), can have eight possible values $([0,0,0],...,[1,1,1])$. Therefore, according to Equation~\ref{eq:objective-function-2}, we have that:

    \begin{align*}
        O([0,0,0]) = 0 &, \qquad O([0,0,1]) = -0.162\\
        O([0,1,0]) = -0.131 &, \qquad O([0,1,1]) = -0.279\\
        O([1,0,0]) = -0.113 &, \qquad O([1,0,1]) = -0.208\\
        O([1,1,0]) =  -0.259 &, \qquad O([1,1,1]) = -0.265\\
    \end{align*}

    Hence, the choice of two data points (index 4,7) yields the optimal result.
\end{example}

In the final step of the process (Phase~3, where we mutate the trajectory), we modify the area surrounding each data point selected by the quantum annealer. Hence, if two selected data points in the solution were close to each other, the mutated area may overlap, reducing the effectiveness of mutation. To address this issue, we place a constraint in the minimum distance between data points in a solution. Specifically, for each pair of variables $x_i$ and $x_{i+1}$, they cannot be selected simultaneously, meaning $x_i$ and $x_{i+1}$ cannot both be set to 1 at the same time if they are within a minimum temporal distance of 2 seconds. Since the objective function is a minimisation problem, we introduce a large penalty value $P$ and multiply it by each pair of $x_ix_{i+1}$.

\begin{equation}
    \min C(\boldsymbol{x}) = P \cdot x_ix_{i+1} 
\end{equation}

Finally, the overall objective function is shown below. We can merge all the linear terms and coefficient terms to create the overall QUBO formulation.

\begin{equation}
    \min Obj(\boldsymbol{x}) = \sum_{mr\in\{cr, id, od\}}w_{mr}\mathit{Obj}^{\mathsf{mr}}(\boldsymbol{x})+w_{num}\mathit{Obj}^{\mathsf{num}}+C(\boldsymbol{x})
\end{equation}

Once all of the sub-solutions have been found, they are merged to compose the final problem and given to the quantum annealer. In summary, we quantum annealer finds the optimal data points for each sub-problem and then finds the optimal data points out of those to determine the final solution.

\subsection{Phase~3: Mutate}
\label{sec:phase3}

In this phase, we first generate the mutations, and then we embed them into the existing test cases.

\emph{\textbf{Generating mutations}:} The purpose of mutation is to potentially enhance the effectiveness in the surrounding of identified data points. Using effectiveness metrics from Phase~1, mutations are over the selected data points from Phase~2. For each identified data point $I(t)$, we define a surrounding temporal range and calculate the correlation value $c(t)$ between the data point value $I(t)$ and the corresponding effectiveness value $\mathcal{E}(t)$. The correlation values are normalized between $[-1,1]$, which are used to define the mutated value of the identified point:
\[I(t)_{mut} = I(t)^{(1-c(t))} * (\mathcal{R}_{max}-\mathcal{R}_{min}) + \mathcal{R}_{min}\]
In this formula, $\mathcal{R}_{max}$ and $\mathcal{R}_{min}$ represent the highest and lowest values. Positive $c(t)$ biases the generation towards $\mathcal{R}_{max}$, favouring an increase, while negative biases it towards $\mathcal{R}_{min}$, favouring a decrease. The intensity of this bias correlates with the magnitude of the normalised value.

\begin{figure}[!h]
	\centering
	\includegraphics[width=0.5\textwidth]{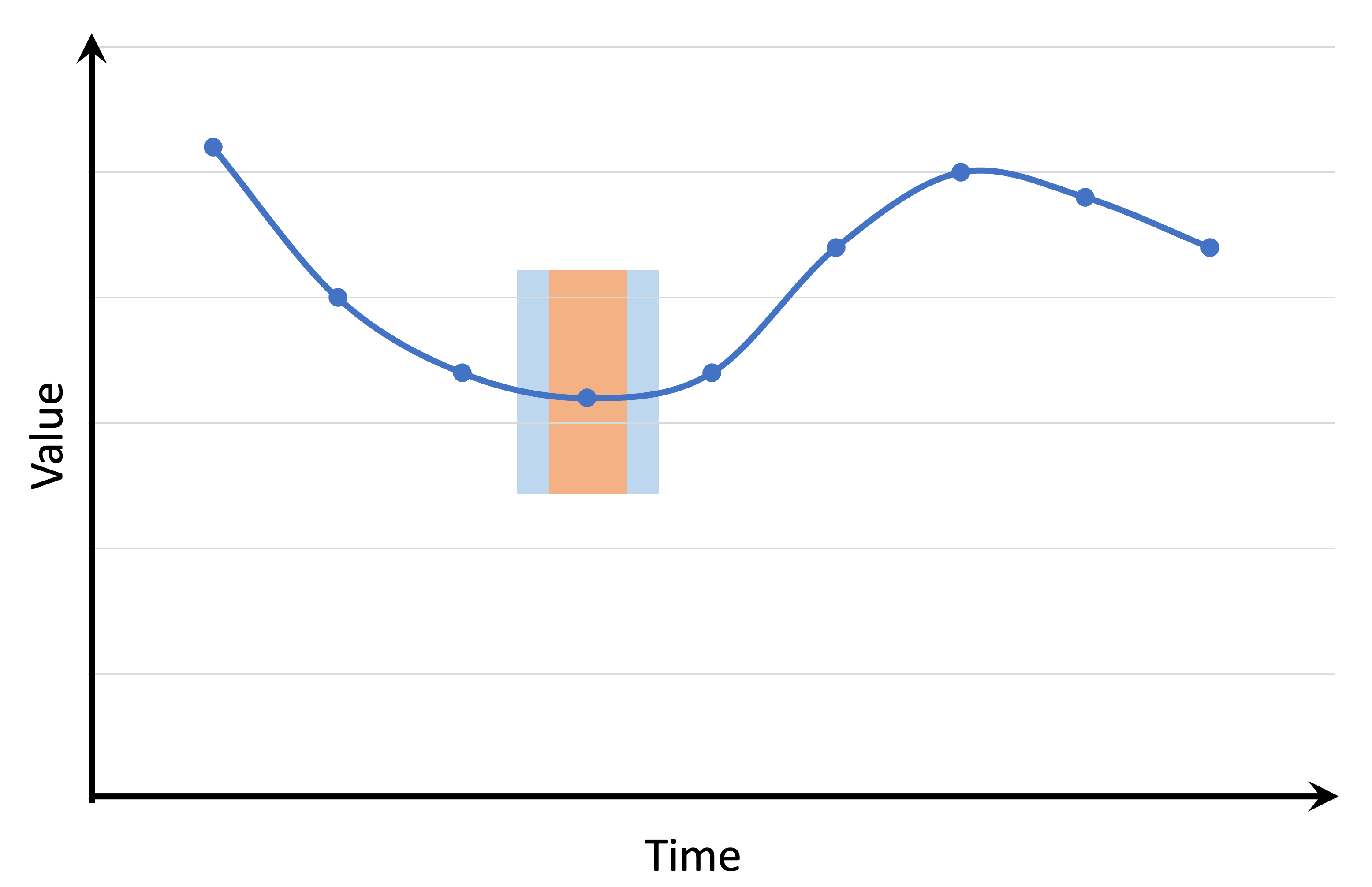}
	\caption{Target area of a test case mutation.}
	\label{fig:mutated}
\end{figure}

\emph{\textbf{Embedding mutations:}} Once mutations are generated, they are embedded within test cases by selecting a region surrounding each mutation. This adjacent region, shown in green in Figure~\ref{fig:mutated} is modified to connect the existing input trajectory with the mutated trajectory. This is achieved using a curve-fitting linear regression technique, similar to the one used in Phase~1.

The output of Phase 3 is an enriched test suite with additional test cases that are modified to target and enhance fault detection effectiveness through mutation-driven adaptations.

\begin{example}

Consider our running example. Figure~\ref{fig:mutated-test-case} depicts the mutated input without (Figure~\ref{fig:re-mutated-input-no-smoothing}) and with curve fitting (Figure~\ref{fig:re-mutated-input-smoothing}). The respective outputs are also shown in this figure (Figures~\ref{fig:re-mutated-output-no-smoothing} and~\ref{fig:re-mutated-output-smoothing}, respectively). Note how the distance between the expected and observed behaviour is more prominent compared to when the non-mutated input was used (Figure~\ref{fig:re-ot3}), which makes faults easier to detect.

    \begin{figure}[!h]
  \centering
    \begin{subfigure}{0.42\linewidth}
      \centering
      \includegraphics[width=1\linewidth]{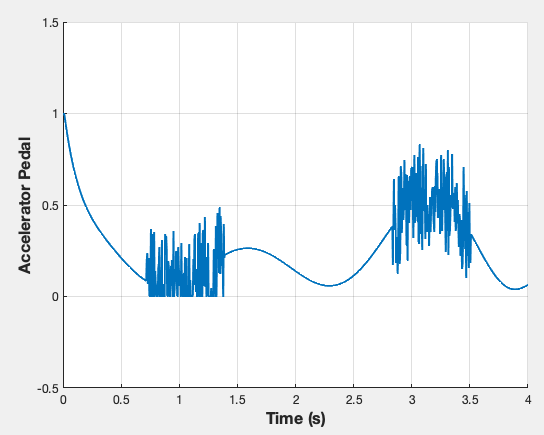} 
      \caption{Mutated input without curve fitting.}
      \label{fig:re-mutated-input-no-smoothing}
    \end{subfigure} \qquad
    \begin{subfigure}{0.42\linewidth}
      \centering
      \includegraphics[width=1\linewidth]{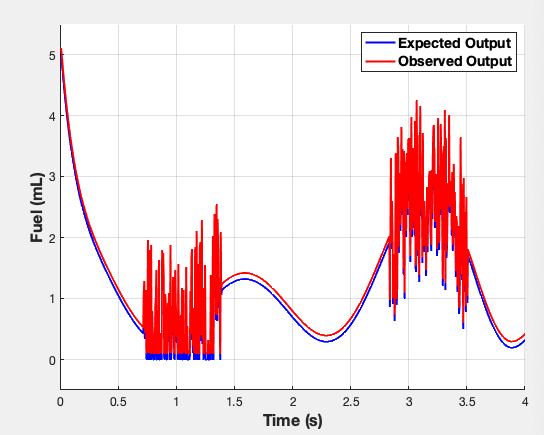} 
      \caption{Output of input without curve fitting.}
      \label{fig:re-mutated-output-no-smoothing}
    \end{subfigure} \\
    \begin{subfigure}{0.42\linewidth}
      \centering
      \includegraphics[width=1\linewidth]{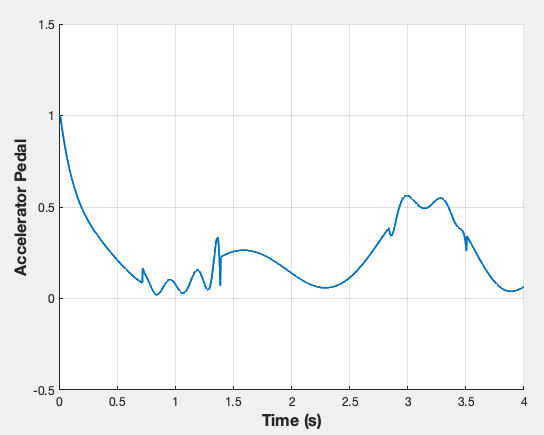}
      \caption{Mutated input with curve fitting.}
      \label{fig:re-mutated-input-smoothing}
    \end{subfigure} \qquad
    \begin{subfigure}{0.42\linewidth}
      \centering
      \includegraphics[width=1\linewidth]{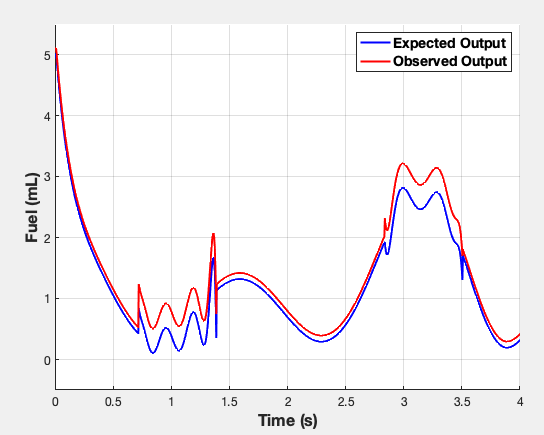} 
      \caption{Outputs of input with curve fitting.}
      \label{fig:re-mutated-output-smoothing}
    \end{subfigure} \hfill
    \caption{Mutated Test Case 3 and its outputs.}
  \label{fig:mutated-test-case}
\end{figure}
    
\end{example}

\section{Controlled Experiment}
\label{sec:experiment}

In this section, we present the application of the strategy proposed in Section~\ref{sec:process} to two case studies: an autonomous vehicle platoon and a pneumatic suspension system. We first outline our research objectives (Section~\ref{sec:research-questions}), then we provide a description of the case studies (Section~\ref{sec:case-studies}), followed by the methodology definition (Section~\ref{sec:methodology}) and lastly, the results and discussion (Section~\ref{sec:results}).

In our methodology, we use system specifications developed in Simulink~\cite{documentationsimulation} - a widely-used development and simulation language for CPS - as well as variants in which faults have been manually inserted. We generate test suites using our process (described in Section~\ref{sec:process}) and use their outputs to reach a verdict on how many of the inserted faults have been detected. The verdict is given by a conformance notion that, for the same input, checks that the distance between the outputs from the correct and faulty models is not greater than a maximum threshold $\epsilon$. 

\subsection{Research objectives}
\label{sec:research-questions}

The main goal of our evaluation is to explore (in terms of efficiency and effectiveness) the use of quantum annealing for test case generation. To this end, we consider two case studies that manifest safety-critical hazards (such as vehicular collisions) that require rigorous verification. We first explore the use of quantum annealing across different problem sizes and number of sub-problems to find optimal configuration values; we then use those values to reach our conclusions on the effectiveness of the generated test suite and the efficiency to generate test suites. Based on this, we aim to answer the following research questions.

\begin{itemize}

    \item \textbf{RQ1:} When using quantum annealing, what is the correlation between problem size, the number of physical qubits, and effectiveness of the results?

    \item \textbf{RQ2:} Is using quantum annealing to mutate the test cases more effective compared to the alternatives?

    \item \textbf{RQ3:} Is using quantum annealing to mutate the test cases more efficient compared to the alternatives, based on experimental results and theoretical analysis?
    

\end{itemize}

The above research questions aim to assess if our quantum annealing approach can lead to effectively and efficiently identifying common types of faults. In RQ1, we explore how different problem sizes affect the quantum annealing process and its results. We conduct an investigation into optimising the number of qubits and their correlation with the effectiveness of the results. The optimal values found via RQ1 are used in the remaining research questions. Effectiveness (RQ2) is measured by comparing the percentage of the inserted faults that the resulting test suite has identified. Efficiency (RQ3) is determined by measuring the amount of time it takes to generate the test suite according to experimental results. We also conducted a theoretical time complexity analysis to compare our approach and alternatives. 

\subsubsection{Baselines}

We compare our approach against test suites generated from two state-of-the-art alternative heuristics, as well as a random test case generation approach. As one of the baseline approaches, we use the classical counterpart to quantum annealing (namely, simulated annealing~\cite{bertsimas1993simulated}) and,  additionally, as the second baseline, we use NSGA-II~\cite{deb2002fast}. We have used the D-Wave API~\footnote{\url{https://docs.ocean.dwavesys.com/en/stable/docs_neal/reference/sampler.html}} to implement the simulated annealing approach and the Pymoo API~\cite{pymoo} to implement the NSGA-II approach. Both heuristics are used to solve the same optimisation problem from Phase~2. In both the simulated annealing and the NSGA-II cases, we employ a single function that combines multiple objectives as one objective  
($\min f(x)=w_1f_1(x) + ... + w_nf_n(x)$) where we have explored different weights to approximate a Pareto front.

Lastly, as a third alternative approach, we randomly select areas of the inputs to be modified. The first two approaches are suitable alternatives to quantum annealing, and the latter is akin to standard mutation-based fuzzing~\cite{miller2007analysis}. 

\subsection{Case Studies}
\label{sec:case-studies}

In this section, we describe the two case studies used throughout our empirical evaluation. 

\subsubsection{Platoon of connected vehicles}
\label{sec:case-study-platooning}

\begin{figure}[h]
	\centering
	\includegraphics[width=0.90\textwidth]{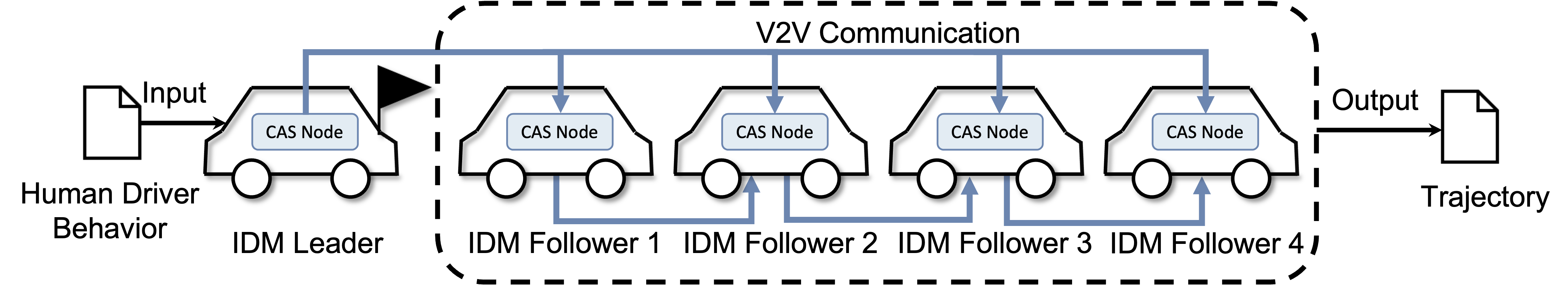}
	\caption{Vehicle platooning \cite{lyamin2017cooperative}.}
	\label{fig:platooning}
\end{figure}

Vehicular platooning is a cooperative and autonomous driving technology that connects two or more vehicles into a convoy. The primary objective of the convoy is to maintain a close yet safe distance between vehicles using vehicle-to-vehicle (V2V) communication and automated driving technologies~\cite{bergenhem2012overview}. This study utilises an open-source model of the system, originally developed for a software verification study~\cite{araujo2020connected}. The system aims to ensure that all follower vehicles in the platoon stay behind the lead vehicle, maintain safe distances from each other, and remain within communication range, as illustrated in Figure \ref{fig:platooning}.

To enable follower vehicles to accelerate and decelerate automatically, an Intelligent Driver Model (IDM) controller~\cite{treiber2000congested} is employed. Each vehicle is also equipped with a communication node responsible for transmitting data packets. The model focuses solely on longitudinal vehicle movement, assuming the platoon operates along a straight, extended highway without significant directional changes. The main input for the testing campaign is the behaviour of the human driver in the lead vehicle, specifically their acceleration pattern. Once this input is provided, V2V communication ensures that the follower vehicles autonomously adjust their behaviour to align with the lead vehicle. The system's output is a trajectory that captures the acceleration of the follower vehicles.



\subsubsection{Suspension System}

\begin{figure}[!h]
	\centering
    \begin{subfigure}[b]{0.45\textwidth}
        \centering
    	\includegraphics[width=130pt]{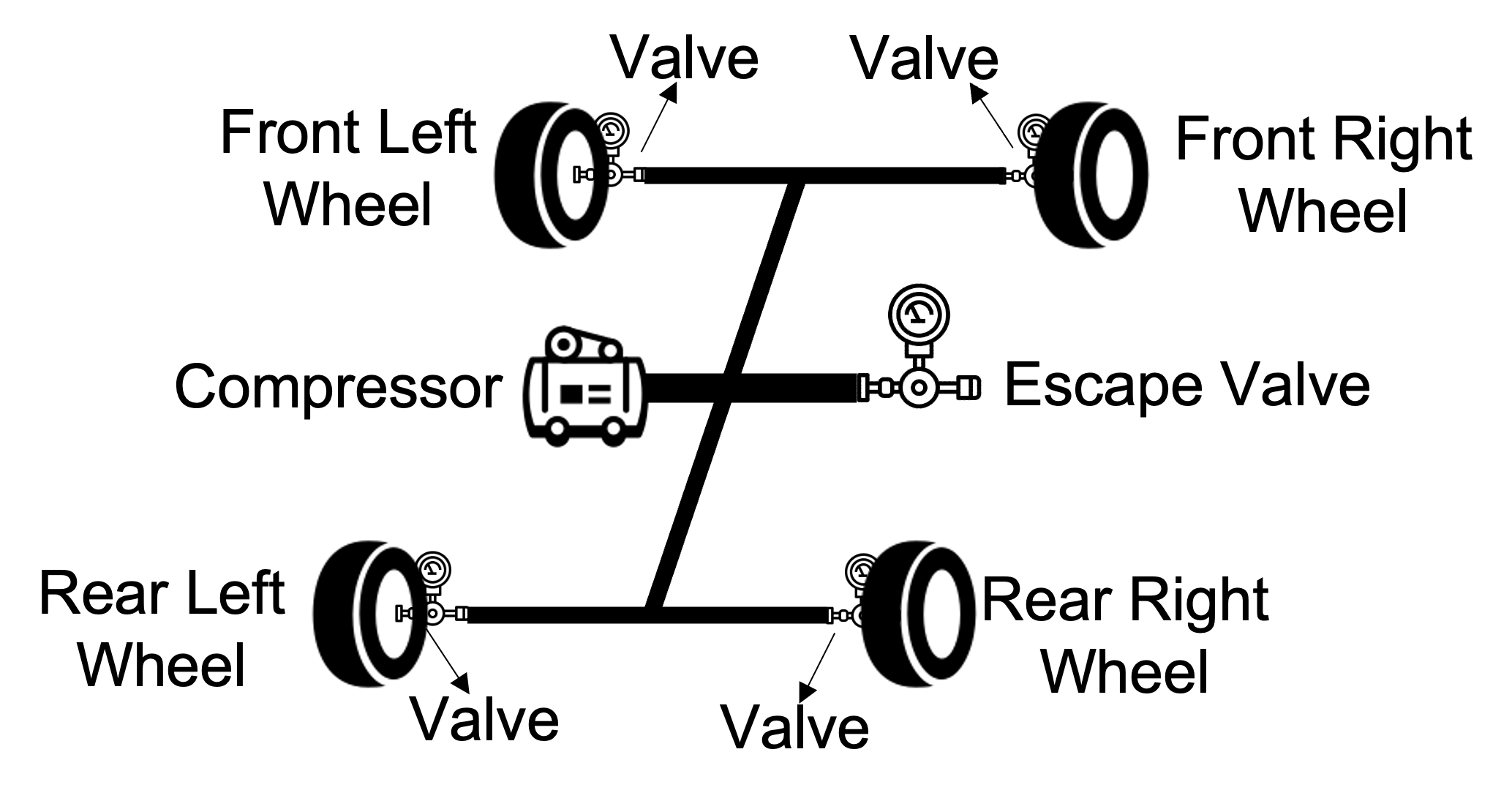}
        \caption{Suspension system overview.}
    	\label{fig:suspension}
    \end{subfigure}
    \hspace*{0.1cm}
    \begin{subfigure}[b]{0.45\textwidth}
        \centering
    	\includegraphics[height=100pt]{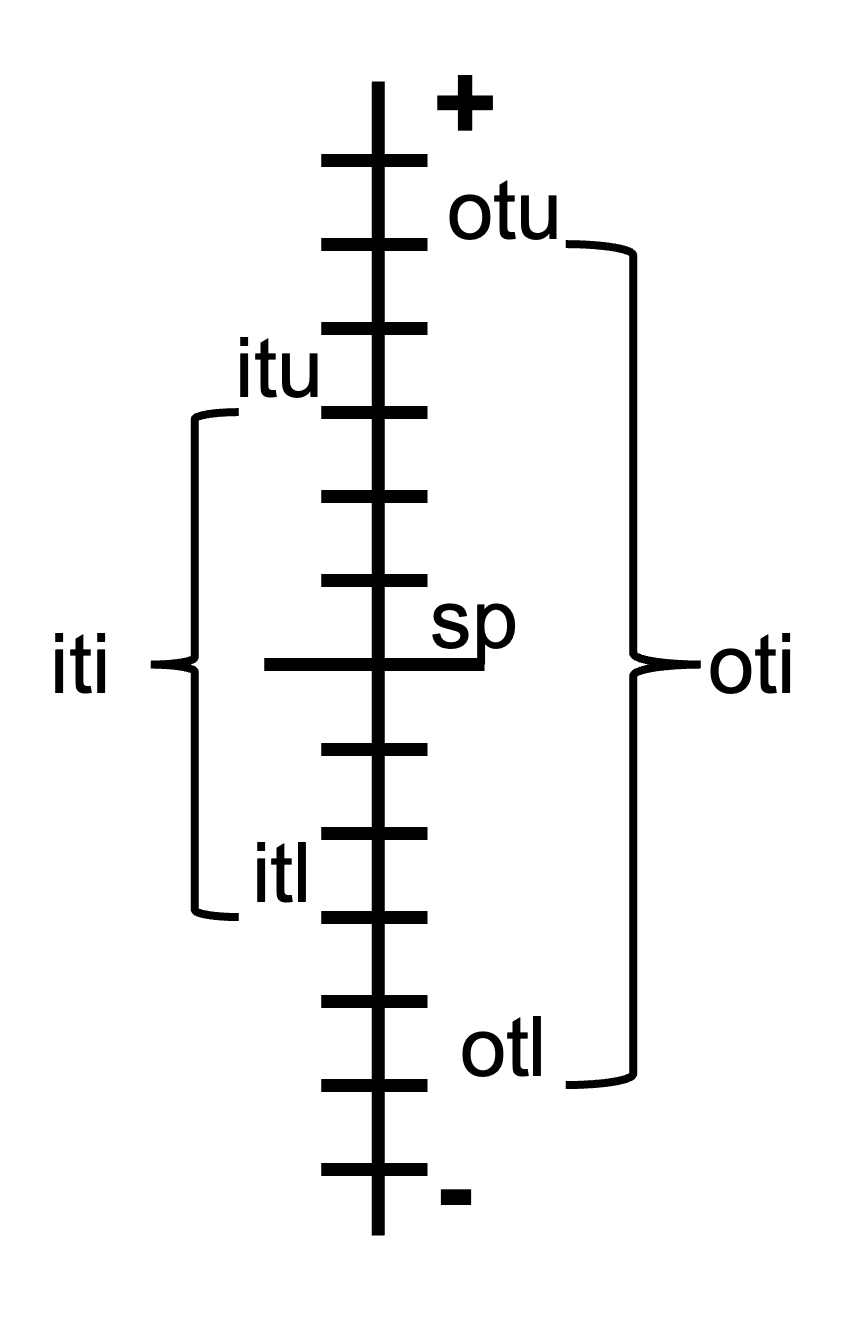}
        \caption{Tolerance levels.}
    	\label{fig:suspensionTolerance}
	\end{subfigure}
  	\caption{Suspension system.}
    \label{fig:suspensionSystem}
\end{figure}

In this case study, we analyse an automotive pneumatic suspension system \cite{muller2000modelling}, designed to enhance driving comfort by maintaining the chassis level and compensating for road irregularities. The system achieves this by utilising a suspension mechanism that connects valves at each wheel to a central compressor and an escape valve (see Figure~\ref{fig:suspension}).

The primary objective is to maintain the chassis level as close as possible to a pre-defined set point for each wheel. Adjustments are guided by tolerance intervals specific to each wheel, as illustrated in Figure~\ref{fig:suspensionTolerance}. These intervals are defined as $[sp - otl, sp + otu]$ for the outer range and $[sp - itl, sp + itu]$ for the inner range, where $sp$ denotes the set point, or the target chassis level. The variables \emph{itl, itu, otl}, and \emph{otu} represent the lower and upper thresholds for the inner and outer tolerance intervals.

The system processes four inputs: (i) road disturbance levels, (ii) adjustments to the chassis level made by the compressor, (iii) adjustments made by the escape valve, and (iv) an indicator signalling whether the vehicle is turning, during which, pneumatic adjustments are restricted. The system's output is the current chassis level, which reflects its continuous effort to stabilise and optimise ride comfort.

\subsection{Experiment Design}
\label{sec:experiment-design}

In this section, we explain the experiment design. Particularly, we describe the methodology (Section~\ref{sec:methodology}), and our metrics and hypotheses (Section~\ref{sec:metrics}).




\subsubsection{Methodology}
\label{sec:methodology}

An overview of the experiment methodology is as follows. First, we insert faults into our extensively tested and (believed-to-be) correct system model. This is done by inserting one fault in the model at a time and, thus, generating a faulty variant. Then, we employ quantum annealing as well as the three alternatives to generate the test suites. More specifically, the decision of which regions of the test cases to be mutated (i.e., Phase~2) is done by four heuristics: (i) quantum annealing, (ii) simulated annealing, (iii) NSGA-II, and (iv) a random-based decision. To obtain statistical significance, the test suite generation and execution parts of the experiment are repeated 10 times, followed by a statistical analysis of the results which is used to answer the RQs. 

\emph{\textbf{Generating faulty variants:}} We employ the Simulink mutation tool FIM~\cite{bartocci2022fim} in our method to insert faults. We insert five types of faults: the \emph{Delay Operator} simulates the introduction of delays, while the \emph{Noise Operator} adds noise to signals within a Simulink model. The \emph{Package Drop} operator replaces a variable's value with a different one. Additionally, the \emph{Logical Operator Replacement} and \emph{Arithmetic Operator Replacement} mutators substitute an operator with another of the same type. These mutation operators were selected based on their applicability to our models. In total, we introduced 10 faults of each type into each case study. We avoided inserting multiple faults simultaneously, as this could lead to unpredictable interactions, such as one modification cancelling out the effect of another.

\emph{\textbf{Generating test suites:}} Next, we generate the test suites: an initial test suite (i..e, obtained from Phase~1) and four mutated ones through quantum annealing and the baseline alternatives. 

To answer RQ1, we compare the results of varying configurations of quantum annealing versus different problem sizes and number of sub-problems. We first investigate the effect of varying the problem sizes (i.e., the number of data-points given to the annealer). Then, we investigate how different numbers of sub-problems can affect the results, by splitting the test case into a varying number of sub-regions, from one all the way up to ten sub-problems.

To answer RQ2 and RQ3, we compare the results between the test suite generated from the quantum annealing approach versus the ones generated by the alternative approaches. When using quantum annealing, we have made use of the results from RQ1 to configure the optimal problem size and number of sub-problems. When using the alternative approaches, unlike with quantum annealing, there's no restriction to the number of variables and, hence, we do not divide the problem into sub-problems or make use of sampling. Instead, the entire trajectory is used by the baselines to decide on the optimal candidates. 

To remove bias, we use the same weights and objectives (namely, maximising criticality, minimising input and output diversity and the number of points selected) as well as the proximity constraint. The weights are set to $0.25$ for criticality, $0.125$ for input diversity,  $0.125$ for output diversity, and $0.5$ for number of data-points. In the case of quantum annealing, the weight for the proximity constraint is set to $1000$\footnote{As explained in Section~\ref{fig:process}, constraints are added to the formula by setting a really high weight.}. The above choice of weights is based on a set of pilot experiments to determine these values. Lastly, we restricted the size of the new test suite to 50 test cases. 



\emph{\textbf{Gathering the results:}} We execute the test suites on our systems to collect the outputs, which are given to a test oracle to determine the verdicts.  More specifically, given one of the test suites, we execute it on the correct model and on a faulty variant and check whether the fault is detected by comparing the outputs. We execute this step for each combination of (test suite $\times$ faulty variant) to calculate how many faults each test suite detects. As is the case with test cases, the outputs of this execution are also represented by trajectories. Thus, input and output trajectories capture the history of input and output variables throughout the execution of the system, respectively. We say that a test suite has identified an inserted fault if, when executed on the system where that fault was inserted, one of its test cases yields a fail verdict. 



\subsubsection{Metrics and Hypotheses}
\label{sec:metrics}
\label{sec:hypotheses}

In this experiment, we hypothesise about the correlation between problem size and number of qubits and, additionally, about the efficacy and efficiency of quantum annealing. In terms of metrics, we collect the percentage of faults detected (PFD) by each approach as well as the time taken to generate the test suite. These metrics essentially quantify effectiveness in terms of the fault detection rate and effectiveness in terms of test case generation time. We have defined one hypothesis for each of our research questions and they are presented in Table~\ref{tab:hypotheses}.

\setlength{\tabcolsep}{1pt}
\begin{table}[!h]
\centering
\caption{Experiment hypotheses.}
\label{tab:hypotheses}
\begin{tabular}{|lc|c|lc|c|lc|}
\hline

\multicolumn{2}{|c|}{\textbf{Hypothesis A}} & & 
\multicolumn{2}{|c|}{\textbf{Hypothesis B}} & & \multicolumn{2}{|c|}{\textbf{Hypothesis C}} \\ \hline

\multicolumn{1}{|l|}{$\ H_{A0}\ $} & $f(p)$ is linear & & \multicolumn{1}{|l|}{$\ H_{B0}\ $} & $\ PFD_{qt} \leq PFD_{alt}\ $ & & \multicolumn{1}{|l|}{$\ H_{C0}\ $} & $\ Time_{qt} \geq Time_{alt}\ $ \\ \hline

\multicolumn{1}{|l|}{$\ H_{A1}\ $} & $f(p)$ is non-linear & & \multicolumn{1}{|l|}{$\ H_{B1}\ $} & $\ PFD_{qt} > PFD_{alt}\ $ & & \multicolumn{1}{|l|}{$\ H_{C1}\ $} & $\ Time_{qt} < Time_{alt}\ $ \\ \hline
\end{tabular}
\end{table}

For RQ1, where we hypothesise about the correlation between the number of qubits and the problem size, we have devised Hypothesis A; the null hypothesis ($H_{A0}$) states that the number of physical qubits (given by $f(p)$) employed by the quantum annealer grows linearly with the problem size ($p$). This experiment aims to refute such a hypothesis. Thus, an alternative hypothesis ($H_{A1}$) is also defined, which can be accepted in case its counterpart is rejected.

With respect to RQ2, which focuses on improvements made by the process with quantum annealing, we compare the fault detection rate of the quantum annealing test suite ($PFD_{qt}$) against that of the alternative test suites ($PFD_{alt}$). We hypothesise that that quantum annealer is more effective than the alternatives and, hence, we also aim to refute the null hypothesis. Analogously, to answer RQ3, we have defined Hypotheses C, which compares the time taken to generate the test cases. 

\subsection{Results}
\label{sec:results}

In this section, we present and discuss the results of our experiment. We answer each research question by providing concrete data and offering insights about the use of quantum annealing. A brief conclusion of our findings can be found in Section~\ref{sec:summary}.

\subsubsection{RQ1: When using quantum annealing, what is the correlation between problem size, the number of physical qubits, and effectiveness of the results?}

To address this question, we executed the quantum annealer with varying numbers of data points, collected the metrics, and generated mutated test suites as outlined in our process. These suites were then executed against faulty variants of our case studies to compute fault detection rates. Table~\ref{tab:results-rq1} summarises the results and displays the average number of physical qubits (NQ) employed, the final fitness function value (FV), the fault detection rate (FD) of the test suite, and the time taken in seconds. Importantly, to collect these metrics, the fault detection rates were computed using data points from the entire trajectory without dividing the problems into sub-problems.

\begin{table}[!h]
\centering
\begin{tabular}{|c|cccc|cccc|}
\hline
 & \multicolumn{4}{c|}{\textbf{Suspension}} & \multicolumn{4}{c|}{\textbf{Platooning}} \\ \cline{2-9} 
\multirow{-2}{*}{\textbf{Problem Size}} & \multicolumn{1}{c|}{\textbf{NQ}} & \multicolumn{1}{c|}{\textbf{FV}} & \multicolumn{1}{c|}{\textbf{FD}} & \textbf{Time} & \multicolumn{1}{c|}{\textbf{NQ}} & \multicolumn{1}{c|}{\textbf{FV}} & \multicolumn{1}{c|}{\textbf{FD}} & \textbf{Time} \\ \hline
5 & 11.7 & 0.2885 & 54.2 & 0.19 & 13.5 & 0.3185 & 48.3 & 0.19 \\
10 & 18.1 & 0.2997 & 54.4 & 0.29 & 19.4 & 0.3342 & 49.5 & 0.37 \\
15 & 33.3 & 0.3212 & 56.6 & 0.39 & 29.5 & 0.3271 & 46.1 & 0.69 \\
20 & 54.8 & 0.2148 & 58.7 & 0.64 & 53.9 & 0.3109 & 47.3 & 0.84 \\
25 & 92.3 & 0.2113 & 56.3 & 0.68 & 91.1 & 0.3055 & 41.4 & 1.19 \\
30 & 125.9 & 0.2105 & 61.0 & 0.99 & 127.0 & 0.2980 & 49.8 & 1.26 \\
35 & \cellcolor[HTML]{9AFF99}160.4 & \cellcolor[HTML]{9AFF99}0.2013 & \cellcolor[HTML]{9AFF99} 64.2 & \multicolumn{1}{l|}{\cellcolor[HTML]{9AFF99}1.01} & 160.0 & 0.3011 &  52.3 & 1.49 \\
40 & 205.3 & 0.2198 & 68.9 & 1.19 & \cellcolor[HTML]{9AFF99}198.2 & \cellcolor[HTML]{9AFF99}0.2901 & \cellcolor[HTML]{9AFF99} 52.1 & \cellcolor[HTML]{9AFF99}1.86 \\
45 & 287.8 & 0.2476 & 69.7 & 1.36 & 281.4 & 0.3683 & 52.0 & 2.08 \\
50 & 326.7 & 0.3124 & 57.8 & 1.59 & 328.9 & 0.3792 & 51.9 & 2.22 \\
60 & 465.1 & 0.3075 & 56.3 & 1.78 & 466.2 & 0.3772 & 51.9 & 2.79 \\
70 & 649.6 & 0.3689 & 58.0 & 2.10 & 637.9 & 0.3801 & 50.8 & 3.15 \\
80 & 912.4 & 0.3591 & 53.2 & 2.47 & 922.5 & 0.3792 & 51.5 & 3.26 \\
90 & 1025.6 & 0.3871 & 56.9 & 2.67 & 1046.8 & 0.3859 & 52.5 & 4.09 \\
100 & 1282.7 & 0.3982 & 65.2 & 3.01 & 1271.2 & 0.3878 & 53.9 & 4.57 \\ \hline
\end{tabular}
\caption{RQ1 Results}
\label{tab:results-rq1}
\end{table}

We define the problem size that yields the lowest fitness function value as the best configuration. Table~\ref{tab:results-rq1} indicates optimal problem sizes of 35 and 40 for the suspension and platooning systems, respectively. Consequently, we selected 40 as the default problem size for subsequent experiments.

To further refine the approach, we analysed the effect of dividing the problem into different numbers of sub-problems. Using a fixed problem size of 40, we split the input trajectories into varying number of slices, ran the quantum annealer on each slice, and then merged the sub-solutions to form the final solution.

\begin{figure}[!h]
	\centering
    \begin{subfigure}[b]{0.45\textwidth}
        \centering
    	\includegraphics[width=1\textwidth]{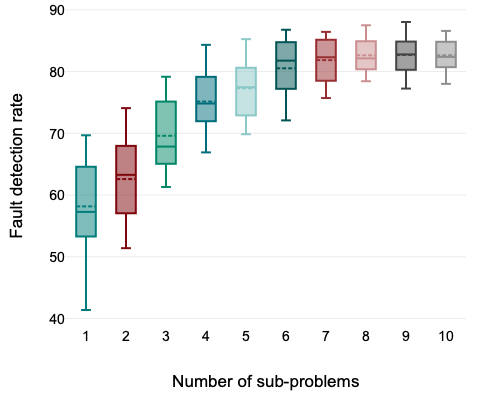}
        \caption{Suspension system.}
    	\label{fig:results-rq1-ss}
    \end{subfigure}
    \hspace*{\fill}
    \begin{subfigure}[b]{0.45\textwidth}
        \centering
    	\includegraphics[width=1\textwidth]{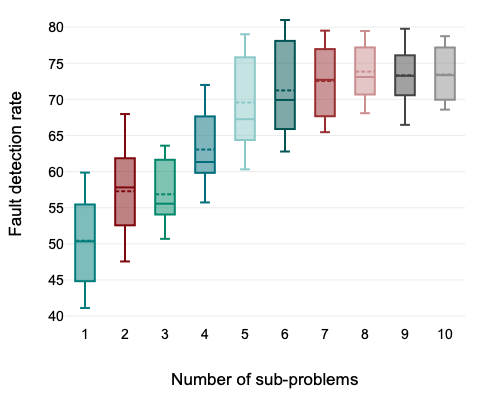}
        \caption{Platooning.}
    	\label{fig:results-rq1-pl}
	\end{subfigure}
  	\caption{Finding optimal number of sub-problems.}
    \label{fig:results-rq1}
\end{figure}

Results, depicted in Figure~\ref{fig:results-rq1}, revealed consistent patterns across both case studies: fault detection rates improved with increasing sub-problems, plateauing around six sub-problems. Beyond this point, the results exhibited reduced variance, indicating greater consistency. However, increasing the number of sub-problems incurs additional computational overhead due to repeated quantum annealer runs. Based on these observations, we select 8 sub-problems as the optimal number that balances effectiveness and efficiency for all future executions of this experiment. As shown in Figure~\ref{fig:results-rq1}, 8 sub-problems is when the curve hits a plateau and we observe negligible effectiveness gains but lower efficiency.

\subsubsection{RQ2: Is using quantum annealing to mutate the test cases more effective compared to the alternatives?}

\begin{figure}[!h]
	\centering
	\includegraphics[width=0.6\textwidth]{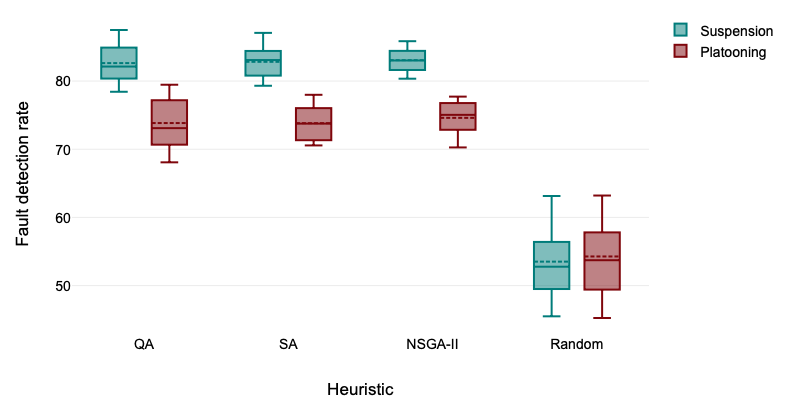}
	\caption{Fault detection rate by heuristic.}
	\label{fig:results-rq2}
\end{figure}

To evaluate the effectiveness of quantum annealing (QA), we compared its fault detection rates against those achieved by simulated annealing (SA), NSGA-II, and random selection using the optimal configurations established in RQ1. Figure~\ref{fig:results-rq2} presents the results for both case studies.

The median fault detection rates for all three heuristics were similar. For the suspension system, quantum annealing achieved a median fault detection rate of $82.6\%$, compared to $82.8\%$ for simulated annealing and $83.0\%$ for NSGA-II. For the platooning system, the respective rates were $73.0\%$, $73.7\%$, and $74.2\%$. No statistically significant differences ($p > 0.05$) were observed among the heuristics, although simulated annealing and NSGA-II exhibited lower variance, likely due to quantum annealing's challenges with noise and embedding. Random selection performed significantly worse than the heuristic approaches, underscoring the value of structured optimisation techniques.

\subsubsection{RQ3: Is using quantum annealing to mutate the test cases more efficient compared to the alternatives?}

Considering that our strategy for using quantum annealing involves dividing the problem into sub-problems and merging them at the end, we have taken this time into account on the data below. Hence, efficiency comparisons considered the time required for decomposition, sampling, and merging, as well as the execution times for all heuristics. For quantum annealing, we utilized D-Wave's API to measure \emph{QPU Access} time\footnote{\url{https://docs.dwavesys.com/docs/latest/c_qpu_timing.html}}, which includes initialisation, programming, and sampling phases (see Section~\ref{sec:quantum-annealing}). We also accounted for the time taken for embedding, a process that, while initially time-consuming, can be reused to achieve substantial speed-ups in subsequent runs.

\begin{table}[!h]
\centering
\caption{Time taken per approach.}
\label{tab:results-rq3}
\begin{tabular}{c|ccc|c|c|c|}
\cline{2-7}
 & \multicolumn{3}{c|}{\textbf{Quantum Annealing}} & \multirow{2}{*}{\textbf{S. Annealing}} & \multirow{2}{*}{\textbf{NSGA-II}} & \multirow{2}{*}{\textbf{Random}} \\ \cline{2-4}
 & \multicolumn{1}{c|}{\textbf{Embedding}} & \multicolumn{1}{c|}{\textbf{Pre and Post}} & \textbf{QPU Access} &  &  &  \\ \hline
\multicolumn{1}{|c|}{\textbf{Suspension}} & 46.56s & \textless 0.2s & 3.57s & 62.3s & 75.2s & \textless 1s \\ \hline
\multicolumn{1}{|c|}{\textbf{Platonning}} & 40.7s & \textless 0.2s & 4.12s & 69.1s & 78.6s & \textless 1s \\ \hline
\end{tabular}
\end{table}

\textbf{Experimental analysis:} The results demonstrate noticeable efficiency gain for quantum annealing. In the suspension system case study, quantum annealing required a median of $3.57$s for QPU access, compared to $62.3$s for simulated annealing and $75.2$s for NSGA-II. In the platooning system, quantum annealing required $4.12$s, compared to $69.1$s and $78.6$s for simulated annealing and NSGA-II, respectively. Random selection was the fastest, taking less than one second in both cases but at the cost of significantly lower fault detection rates. Decomposition and re-composition processes for quantum annealing added approximately $0.2$s seconds per run. Embedding, on the other hand, took between $40$s to $45$s, overshadowing QPU access time. Regardless, even with these additional steps, our approach with quantum annealing remained more efficient overall than the alternatives.

\textbf{Theoretical analysis:} We compare the efficiency of our approach with simulated annealing (i.e., the more efficient baseline approach implemented in the paper) by analysing the time complexity of the two approaches. Let $N$ denote the sub-problem size and $M$ the number of sub-problems. The time complexity of running quantum annealing on a single sub-problem is $\mathcal{O}(e^{\sqrt{N}})$~\cite{mukherjee2015multivariable}, and solving $M$ sub-problems yields the total time complexity of $\mathcal{O}(Me^{\sqrt{N}})$. On the other hand, suppose the number of data points in the entire trajectory is $n$, the time complexity of applying simulated annealing to the full dataset is $\mathcal{O}(e^{n})$~\cite{mukherjee2015multivariable}. Given $M*N<n$ and $M$, $N$, and $n$ are all natural numbers larger than 1, it follows that $Me^{\sqrt{N}}<e^n$ (i.e., since $Me^N$ can be written $e^{\ln{M} + \sqrt{N}}$, the exponent satisfies $\ln{M}+\sqrt{N}<M+N<M*N<n$). This result demonstrates that our approach has a lower time complexity than simulated annealing, showing a higher efficiency of our approach.

\subsubsection{Summary of the results}
\label{sec:summary}

The results of this experiment provide valuable insights into the efficacy and efficiency of quantum annealing for test case generation when compared to state-of-the art heuristics. In RQ1, we demonstrate that quantum annealing strategies are not ready to solve large scale problems without an accompanying decomposition solution. However, even when considering the time required for decomposition and recomposing the problem and solution, quantum annealing maintained superior overall efficiency. As the main drawback in terms of efficiency, embedding takes a substantial amount of time to complete. This can be mitigated by re-using the same embedding for subsequent runs, making quantum annealing a more scalable solution for larger or iterative problems. More efficient techniques for embedding are still the subject of ongoing research~\cite{konz2021embedding, zbinden2020embedding}. In terms of effectiveness, we have observed parity in fault detection rates across heuristics, which demonstrates that no approach is statistically superior in this context. This further highlights quantum annealing as a viable alternative to the state-of-the-art heuristics. 

In summary, these findings underscore the strengths and trade-offs of quantum annealing. It offers clear advantages in computational efficiency and remains competitive in fault detection, but requires careful management of problem decomposition and embedding to minimise variance and ensure consistent results.

\subsection{Threats to Validity}

The following threats to validity have been identified for this experiment.

\begin{itemize}

\item \textbf{Internal Validity:} Our experiment methodology relies on specific types of faults, which might not fully capture the diversity or complexity of faults encountered in real-world CPS applications. In addition, only one fault is inserted at a time to avoid interactions between faults. However, this may not reflect real-world scenarios. To mitigate this issue, we have introduced a total of 50 faults using an external automatic tool found in peer-reviewed literature. As another threat to internal validity, the use of sampling in quantum annealing to handle problem size limitations may introduce bias, as different sampling strategies might yield different results. Investigating the effect of different sampling approaches is outside of the scope of this study and is left for future work. 

\item \textbf{External Validity:} The experiment is conducted on specific Simulink models and, hence, the system under test is a simulation-based model that may lack fidelity. As mitigation to this threat, our models have been developed by domain-experts and have been employed across different experiments available in peer-reviewed literature. As a second threat to external validity, quantum annealing is evaluated under specific problem sizes and sub-problem configurations. Its scalability to even larger problems remains uncertain.

\item \textbf{Construct Validity:} As a threat to construct validity, the quality of the inputs used in this experiments also have an impact on the results. We mitigate this issue by comparing our approach against three baselines used in test case generation. One that generates random but valid inputs and two search-based approaches; it is unclear whether other forms of input generation would have the same effect.

\item \textbf{Conclusion Validity:} As a threat to conclusion validity, this experiment only considers two examples of CPS. This makes it hard to generalise the outcome of this experiment for a general class of cyber-physical systems. This is mitigated by the large number of mutants that were inserted into this system. Moreover, the accuracy of the test oracle, which determines fault detection by comparing model outputs to an expected range, may affect the reliability of results. If the oracle parameters for the euclidean distance do not accurately represent acceptable deviations for all faults, this can affect the  conclusions. To mitigate these issues, we note that the choice of mutation operators and oracles have been made by experts in the field based on domain knowledge and prior, peer-reviewed, experiments.
 
\end{itemize}

\section{Conclusion}
\label{sec:conclusion}

In this work, we introduce a novel approach for efficiently generating effective test cases for Cyber-Physical Systems (CPS) by employing quantum annealing. Our method provides a structured, three-phase process to identify critical regions of test cases based on adequacy metrics, formulate the problem as a combinatorial optimisation using the Quadratic Unconstrained Binary Optimisation (QUBO) model, and apply targeted mutations that enhance fault-detection and test case diversity in the identified regions. Through empirical evaluation on two CPS case studies, we demonstrate that our quantum-based strategy not only improves the effectiveness of a previously generated test suites but also surpasses alternative test case generation methods in terms of efficiency.

Our results affirm that quantum annealing has significant potential to solve complex, optimisation-heavy problems even in test case generation contexts. Although current quantum computing technology has limitations, our findings highlight how it can be adequately utilised within specific problem domains. Future work could explore scaling this approach as quantum hardware and software advance. Moreover, investigating the effects of smarter sampling approaches and different methodologies for mutating the test cases is also part of future research directions.


\bibliographystyle{IEEEtranS}
\bibliography{references}

\end{document}